\documentstyle[11pt,epsf,axodraw]{article}

\def\Year{\expandafter\eatPrefix\the\year}
\newcount\hours \newcount\minutes
\def\monthname{\ifcase\month\or
January\or February\or March\or April\or May\or June\or July\or
August\or September\or October\or November\or December\fi}
\def\shortmonthname{\ifcase\month\or
Jan\or Feb\or Mar\or Apr\or May\or Jun\or Jul\or
Aug\or Sep\or Oct\or Nov\or Dec\fi}

\def\TimeStamp{\hours\the\time\divide\hours by60%
\minutes -\the\time\divide\minutes by60\multiply\minutes by60%
\advance\minutes by\the\time%
${\rm \shortmonthname}\cdot   \if\day<10{}0\fi\the\day\cdot   \the\year%
\qquad\the\hours:\if\minutes<10{}0\fi\the\minutes$}

\newskip\humongous \humongous=0pt plus 1000pt minus 100pt
\def\caja{\mathsurround=0pt}
\def\eqalign#1{\,\vcenter{\openup1\jot \caja
       \ialign{\strut \hfil$\displaystyle{##}$&$
        \displaystyle{{}##}$\hfil\crcr#1\crcr}}\,}
\newif\ifdtup

\newcounter{eqnumber}[section]
\renewcommand{\theeqnumber}{\thesection.\arabic{eqnumber}}
\def\equn{\refstepcounter{eqnumber}
\eqno({\rm \theeqnumber})
}

\def\npb#1#2#3{{ Nucl.\ Phys.\ } {\bf  B#1}:#3 (#2)}
\def\plb#1#2#3{{ Phys.\ Lett.\ } {\bf  B#1}:#3 (#2)}

\def\cqg#1#2#3{{ Class.\ and Quant.\ Grav.} {\bf  #1}:#3 (#2)}

\def\JHEP#1#2#3{{ J.\ High\ Ener.\ Phys.\  } {\bf  #1}:#3 (#2)}

\def\hepth#1{[hep-th/#1]}

\newbox\charbox
\newbox\slabox
\def\s#1{{      
        \setbox\charbox=\hbox{$#1$}
        \setbox\slabox=\hbox{$/$}
        \dimen\charbox=\ht\slabox
        \advance\dimen\charbox by -\dp\slabox
        \advance\dimen\charbox by -\ht\charbox
        \advance\dimen\charbox by \dp\charbox
        \divide\dimen\charbox by 2
        \raise-\dimen\charbox\hbox to \wd\charbox{\hss/\hss}
        \llap{$#1$}
}}

\def\spa#1.#2{\left\langle#1\,#2\right\rangle}
\def\spb#1.#2{\left[#1\,#2\right]}
\def\lor#1.#2{\left(#1\,#2\right)}

\catcode`@=11  

\def\Tr{\, {\rm Tr}}

\def\eps{\epsilon}

\def\pol{\eps}

\def\x#1#2{x_{#1 #2}}

\def\half{{1\over 2}}

\def\lsl{\not{\hbox{\kern-2.3pt $\ell$}}}
\def\ksl{\not{\hbox{\kern-2.3pt $k$}}}

\def\Mloop{M^{\rm 1-loop}}
\def\Floop{F^{\rm 1-loop}}
\def\Mtree{M^{\rm tree}}
\def\CMtree{{\cal M}^{\rm tree}}
\def\CMloop{{\cal M}^{\rm1-loop }}
\def\Mcount{M^{\rm counter}}

\def\Ftree{F^{\rm tree}}

\def\tree{{\rm tree}}

\def\Gbdb{\dot {\overline G}{}_B}
\def\Gbddb{\ddot {\overline G}{}_B}

\def\Gbd{\dot G_B}
\def\Gbdd{\ddot G_B}

\def\sand#1.#2.#3{%
\left\langle\smash{#1}{\vphantom1}^{-}\right|{#2}%
\left|\smash{#3}{\vphantom1}^{-}\right\rangle}
\def\sandp#1.#2.#3{%
\left\langle\smash{#1}{\vphantom1}^{-}\right|{#2}%
\left|\smash{#3}{\vphantom1}^{+}\right\rangle}
\def\sandpp#1.#2.#3{%
\left\langle\smash{#1}{\vphantom1}^{+}\right|{#2}%
\left|\smash{#3}{\vphantom1}^{+}\right\rangle}
\def\sandmm#1.#2.#3{%
\left\langle\smash{#1}{\vphantom1}^{-}\right|{#2}%
\left|\smash{#3}{\vphantom1}^{-}\right\rangle}

\def\beq{\begin{equation}}
\def\eeq{\end{equation}}
\def\beqa{\begin{eqnarray}}
\def\eeqa{\end{eqnarray}}

\font\ninerm=cmr9

\begin{document}

\begin{titlepage}

\begin{flushright}
SWAT-02-366 \\
\end{flushright}

\vskip 2.cm

\begin{center}
\begin{Large}
{\bf Gravity and Form Scattering and 
Renormalisation of Gravity in Six and Eight Dimensions}

\vskip 2.cm

\end{Large}

\vskip 2.cm

{\large 
David C. Dunbar 
and Nicolaus W. P. Turner$^{1}$}

\vskip 0.5cm

{\it Department of Physics, \\
University
    of Wales Swansea, 
\\ Swansea, SA2 8PP, UK }

\vskip 5.0cm

\begin{abstract}

We calculate one-loop scattering amplitudes for gravitons and
two-forms in dimensions greater than four. The string based
Kawai-Lewellen-Tye relationships allow gravitons and two-forms to be
treated in a unified manner.  We use the results to determine the
ultra-violet infinities present in these amplitudes and show how
these determine the
renormalised one-loop action in six and eight dimensions.

\end{abstract}

\vskip 1.0cm

\end{center}

\vfill

\noindent\hrule width 3.6in\hfil\break
%
%
${}^1$Research supported by PPARC \hfil\break

\end{titlepage}

\baselineskip 16pt

















\section{Introduction}

Quantum gravity~\cite{Dewitt} has proven a difficult theory to fit
into the context of quantum field theory.
Due to the dimensionful nature of its coupling constant,
$$
[ \kappa^2 ] = (D-2)
\equn 
$$
any renormalisation of the
theory must involve the introduction of new operators
rather than a redefinition of the coupling constant.  With increasing
loop order increasingly higher dimension operators may appear and we
obtain a theory described by an infinite set of operators which lacks
predictive power. The only escape from such a scenario is if this process
truncates after a finite number of loops and we call such a theory
finite.  The most natural assumption is that additional symmetries
will be needed to forbid the presence of the potential counterterms.
The search for a finite theory has led physicists in many diverse
direction with mostly negative results.  The sole spectacular
candidate of a finite theory, including gravity, lies in superstring
theory~\cite{GSW}.  Although superstring theory is thought to be
finite, the other issue, namely the determination of ultra-violet
infinities in other theories has proved to be a very difficult problem
with few concrete results.  Unless a finite field theory of gravity
can be constructed, gravity must be regarded as a low-energy effective
theory of a more fundamental theory such as string theory. In this
case the low-energy effective action will play the role of the
counterterm action and by studying this we may hope
to learn of the symmetries and properties of the fundamental theory.

In general, in $D$-dimensions, at $L$ loops counterterms such as
$$
\nabla^n R^m
\equn
$$
appear where $n+2m = (D-2)L+2$ and we have suppressed the indices on
the Riemann tensor $R_{abcd}$.  We use forms of dimensional
regularisation to evaluate the ultra-violet structure of a
theory. (And thus only obtaining divergences in even dimensions.)
There are two aspects to determining the counterterms. Firstly one can
determine the possible counterterms consistent with the symmetries and
secondly one must determine their coefficient by specific
calculations.

At one-loop for $D=4$, pure Einstein gravity is actually finite
\cite{OneLoopFinite,Zak}, although matter coupled to gravity is
not~\cite{MatterGravity,DNb}.  Although matter coupled to gravity is
ultra-violet divergent, the divergences do not appear in one-loop
amplitudes with only external gravitons.  Beyond one-loop it has been
shown that pure
gravity has a two-loop
infinity, as first calculated  by Goroff and
Sagnotti~\cite{GoroffSagnotti} and later Van De Ven
\cite{vandeVen}.
Matter in general does not improve renormalisability,
however, special combinations can lead to cancellation of infinities.
The best understood example of this are
theories with supersymmetry which have much better ultra-violet
properties.
For example, $N=4$ super-Yang-Mills is a finite theory \cite{N4finite}
in $D=4$ and
supergravity theories are two-loop finite~\cite{TwoLoopFinite} in $D=4$.

In this paper we calculate divergences appearing in amplitudes in
dimensions higher than four at one-loop and examine the effect of
matter upon the infinities which appear and examine whether there
exist simplifying combinations of matter. We calculate 
amplitudes
with mixtures of
gravitons and antisymmetric two-forms
and 
we determine the divergences
appearing in physical on-shell amplitudes for which many specialised
calculational techniques exist. String theory via the
relations first written down by Kawai, Lewellen and Tye~\cite{KLT} for
tree amplitudes and later further developed for loop amplitudes
\cite{BDDPR} also allows, in some cases, the relatively easy computation
of amplitudes involving gravity from amplitudes which involve gauge
particles.  (Alternative approaches involve the calculation of
off-shell functions typically with a smaller number of legs.)  We
restrict ourselves to four-point amplitudes thus effectively only
being sensitive to counterterms up to $\partial^n R^4$.   We present
particular helicity amplitudes which exhibit divergences in all (even)
dimensions greater than four, thus indicating one-loop counterterms are
always necessary (in even dimensions).  We also use the divergences to
evaluate the form of the counterterms in $D=6$ and $D=8$.  The $D=6$
one-loop result has been previously calculated as a precursor to
calculating the two-loop $D=4$ infinity since both of these have the
same $R^3$ structure. In $D=8$ we have evaluated the exact counterterm
structure for comparison to that found in supersymmetric theories.
For matter coupled gravity, the amplitudes with only external
gravitons do not completely determine the counterterms which
depend exclusively on the Riemann tensor and so we also evaluate
amplitudes which are mixtures of gravitons and antisymmetric two-forms
to enable us to fix the counterterms containing the Ricci tensor.

\section{Organisation of the Amplitudes}

\subsection{Basic Theory}

We consider the calculation of amplitudes with gravity
minimally coupled to a variety of matter. For example
the coupling to a complex scalar and a two-form
is given by
$$
{\cal S} = \int d^Dx \ \sqrt{|g|} \ \left[
\frac{2}{\kappa^2} \, R + \nabla_a \phi^* \, \nabla^a \phi
+ \frac{1}{6} \, F_{bcd} \, F^{bcd}
\right]
\equn
$$
where
$$
\eqalign{
F_{abc} &= \nabla_{a} B_{bc} + \nabla_{b} B_{ca} + \nabla_{c} B_{ab}
\cr}
\equn
$$
and $B_{ab}$ is the two-form field which is antisymmetric.
The field strength, $F_{abc}$,
is invariant under
$$
B_{ab}
\rightarrow
B_{ab} +\nabla_{a}\eta_{b}-\nabla_{b}\eta_{a}
\equn
$$

We investigate the computation of scattering amplitudes in this
theory focusing upon four-point on-shell one-loop
amplitudes.  In a gauge or gravitational theory smaller point
amplitudes vanish on-shell and so the four-point amplitudes are the first
non-trivial amplitudes.  However, as we shall see they contain a
great deal of information regarding the quantum theory.

We calculate amplitudes for dimensions $D > 4$ however, we can simplify
the four-point case  by using the four momenta to define a four
dimensional hyper-plane in $D$ dimensions.  With respect to this
hyper-plane many of the well developed four dimensional
organisational~\cite{ManganoParke,LanceTASI}
techniques can be applied to these calculations.
One of the most useful techniques is that of spinor helicity which,
unfortunately, does not easily generalise to $D> 4$.  However, with
respect to the four dimensional hyper-plane it can still prove a useful
technique which we now describe.

\subsection{$D > 4$ Spinor Helicity }

In four dimensional gauge theory calculations, it is extremely useful
to organise amplitudes according to the  helicity of the
external gluon or quark (or even scalar).

Furthermore one can use  spinor helicity techniques
\cite{SpinorHelicity,ManganoParke}
where the polarisation vector of a gluon is realised as  combinations of
four dimensional Weyl spinors $\vert k^{\pm} \rangle$,
$$
\pol^{+}_{\mu} (k;q) =
{\sandmm{q}.{\gamma_{\mu}}.k
\over  \sqrt2 \spa{q}.k} \hskip 2.0 cm
\pol^{-}_{\mu} (k;q) =
{\sandpp{q}.{\gamma_{\mu}}.k
\over \sqrt{2} \spb{k}.q}
\equn
$$
where $k$ is the gluon momentum and $q$ is an arbitrary null
`reference momentum' which drops out of the final gauge-invariant
amplitudes.  The plus and minus labels on the polarization vectors
refer to the gluon helicities and we use the notation
$\langle ij \rangle\equiv  \langle k_i^{-} \vert k_j^{+} \rangle\, ,
[ij] \equiv \langle k_i^{+} \vert k_j^{-} \rangle$.
These spinor products are anti-symmetric and satisfy
$\spa{i}.j \spb{j}.i = 2 k_i \cdot k_j \equiv s_{ij}$.
For four-point amplitudes we use the usual
Mandelstam variables $s= s_{12}$, $t= s_{14}$ and $u= s_{13}$.

Although spinor helicity is a four dimensional concept it can
be used
in higher dimensions.  First consider the
polarisation tensors for a $D$-dimensional vector particle.  When
considering four-point amplitudes, momentum conservation implies
the first four dimensions can be defined
so that
the momenta of the scattered particles lie
exclusively in this four dimensional hyper-plane.  Defining
$$
x^{a} =( x^{\mu} ; x^I )
\equn
$$
where $x^{\mu}$ denotes the coordinates of the four dimensional hyper-plane
and $x^I$ are the remaining  $(D-4)$.
The coordinates are chosen so
$$
k_i^I =0
\equn
$$
for the four external momenta, $k_i$. Using this frame
we can choose the helicity vectors $\eps_{a}$ to be of
two types: $\eps^{\pm}_{a}$ and $\eps^I_{a}$~\cite{DunbarTurner}
$$
\eqalign{
\eps^{\pm}_{a} &= ( \eps^{\pm}_{\mu} \, ; 0 )
\cr
\eps^I_{a} &= ( \ 0 \ ; 0, \dots ,0,1,0, \dots ,0 )
\cr}
\equn
$$
which provide $(D-2)$ independent polarisation vectors. These satisfy
$$
\eqalign{
\eps^{\pm} \cdot \eps^I = 0 \; ,   \hskip 0.5cm & \hskip 0.5cm
k_i \cdot \eps^I = 0 \; ,  \hskip 0.5cm
\eps^I \cdot \eps^J = -\delta^{IJ}
\cr}
\equn
$$
We use the above polarisations vectors in $D$ dimensions
to construct the graviton
polarisation tensors, which are required to be symmetric, transverse and
traceless.

For the four dimensional case there are only two graviton helicities
whose polarisation tensors can be constructed from direct products of
polarisations vectors~\cite{Berends,GravityHelicity},
$$
\eqalign{
\epsilon_{ab}^{++} & = \epsilon_{a}^+ \, \epsilon_{b}^+
\cr
\epsilon_{ab}^{--} & = \epsilon_{a}^- \, \epsilon_{b}^-
\cr}
\equn
$$
In $D>4$  the additional polarisation tensors may also be constructed from the
polarisation vectors
$$
\eqalign{
\epsilon_{ab}^{+I} & =
\frac{1}{\sqrt{2}} \, \left( \epsilon_{a}^+ \ \epsilon_{b}^I + \epsilon_{a}^I \ \epsilon_{b}^+ \right)
\hskip 4.0 truecm   [D-4]
\cr
\epsilon_{ab}^{-I} & =
\frac{1}{\sqrt{2}} \, \left( \epsilon_{a}^- \ \epsilon_{b}^I + \epsilon_{a}^I \ \epsilon_{b}^- \right)
\hskip 4.0 truecm  [D-4]
\cr
\epsilon_{ab}^{IJ} & =
\frac{1}{\sqrt{2}} \, \left( \epsilon_{a}^I \ \epsilon_{b}^J + \epsilon_{a}^J \ \epsilon_{b}^I \right)
\hskip 4.1 truecm  [(D-4)(D-5)/2]
\cr
\epsilon_{ab}^{II} & =
\sqrt{\frac{2}{3}} \, \left[ \epsilon_{a}^I \ \epsilon_{b}^I - \frac{1}{2} \,
\left( \epsilon_{a}^+ \ \epsilon_{b}^- + \epsilon_{a}^- \ \epsilon_{b}^+ \right) \right]
\hskip 1.8 truecm  [ D-4]\cr}
\equn
$$
where $I \neq J$.
The figures in square brackets refer to the number of independent such polarisations.
Together with $\eps^{++}$ and $\eps^{--}$ they provide the necessary
$(D-2)(D-1)/2-1$ polarisations.

We can also use spinor helicity techniques for the polarisation tensors
of the antisymmetric two-form.  In this case the
polarisation tensors for the two-form, $B_{ab}$, must be transverse and
antisymmetric.
These can also be constructed from the polarisation
vectors
$$
\eqalign{
\epsilon_{ab}^{+-} & =
\frac{1}{\sqrt{2}} \, \left( \epsilon_{a}^+ \ \epsilon_{b}^- - \epsilon_{a}^- \ \epsilon_{b}^+ \right)
\hskip 2.2 truecm  [1] \cr
\epsilon_{ab}^{+I} & =
\frac{1}{\sqrt{2}} \, \left( \epsilon_{a}^+ \ \epsilon_{b}^I - \epsilon_{a}^I \ \epsilon_{b}^+ \right)
\hskip 2.0 truecm  [ D-4] \cr
\epsilon_{ab}^{-I} & =
\frac{1}{\sqrt{2}} \, \left( \epsilon_{a}^- \ \epsilon_{b}^I - \epsilon_{a}^I \ \epsilon_{b}^- \right)
\hskip 2.0 truecm  [ D-4] \cr
\epsilon_{ab}^{IJ} & =
\frac{1}{\sqrt{2}} \, \left( \epsilon_{a}^I \ \epsilon_{b}^J - \epsilon_{a}^J \ \epsilon_{b}^I \right)
\hskip 2.1 truecm  [(D-4)(D-5)/2] \cr}
\equn
$$
providing $(D-2)(D-3)/2$ independent polarisations.

\def\denom{\spa1.2\spa2.3\spa3.4\spa4.1}

\section{Kawai-Lewellen-Tye Relationships}

The Kawai-Lewellen-Tye~(KLT)~\cite{KLT} relationships express closed string
tree amplitudes as sums of products of open string tree amplitudes.
Heuristically there is a very obvious relationship between the amplitudes of
closed and open strings since an open string amplitude may be written~\cite{GSW}
$$
A_{open} \sim \int dx  K
\equn$$
where K is a ``kinematic factor''
and a
closed string amplitude may be written
$$
A_{closed} \sim \int d^2 z  K_l \times K_r
\equn$$
where the $K_l$ and $K_r$ are individually the kinematic factors for a
open string theory.  This heuristic argument suggests a relationship,
however the suggested relationship is weaker than that contained in
the KLT relations. (The proof is far from trivial.) For four and
five-point
amplitudes the KLT-relationship is
$$
\eqalign{
M_4^{\rm tree}( 1,2,3,4)
&= -\frac{i s_{12}}{4} \, A^{\rm tree}_4(1,2,3,4) \, A^{\rm tree}_4(1,2,4,3)
\cr
M_5^{\rm tree}(1,2,3,4,5) &= \
\frac{i s_{12} s_{34}}{8} \, A_5^{\rm tree}(1,2,3,4,5) \, A_5^{\rm tree}(2,1,4,3,5)
\cr
& \hskip 1cm
 + \frac{i s_{13}s_{24}}{8} \, A_5^{\rm tree}(1,3,2,4,5) \, A_5^{\rm tree}(3,1,4,2,5)
\cr}
\equn
$$
where $M_4( 1,2,3,4)$ is a closed string amplitude and $A_4(1,2,3,4)$ are
color-stripped open string partial amplitudes.
These exact relationships between open and closed string tree amplitudes
becomes, in the infinite string tension limit, a relationship
between the field theory amplitudes for
massless particles.

The $M_n$'s are the amplitudes in a gravity theory and
the $A_n$'s are the color-ordered partial amplitudes in a gauge
theory. The full gauge theory amplitude is  obtained by multiplying the
$A_n$ by color-traces~\cite{ManganoParke,BKColor,BDKReview}
$$
\eqalign{
{\cal A}_n^{\rm tree}& (1,2,\ldots ,n) =
 g^{n-2} \sum_{\sigma \in  S_n/Z_n}
{\rm Tr}\left( T^{\alpha_{\sigma(1)}}
\cdots  T^{\alpha_{\sigma(n)}} \right)
 A_n^{\rm tree}(\sigma(1), \ldots, \sigma(n)) \cr} \hskip .6 cm
\equn$$
where $S_n/Z_n$ is the set of all permutations, but with cyclic
rotations removed, and $g$ is the gauge theory coupling constant.
The $T^{\alpha_i}$ are fundamental representation
matrices for the Yang-Mills gauge group $SU(N_c)$, normalized so that
$\Tr(T^\alpha T^\beta) = \delta^{\alpha \beta}$.
For states coupling with the strength of gravity,
the full amplitude including the gravitational coupling constant is,
$$
{\cal M}_n^{\rm tree} (1,\ldots , n) =
\kappa^{n-2}
M_n^{\rm tree}(1,\ldots , n)
\equn
$$

Consider the case where the massless open
string states are vector bosons described by polarisation vectors
$\eps_i$. Then the open string amplitudes will
be
$$
A_4^{\rm tree}(\eps_1,\eps_2,\eps_3,\eps_4)
\equn
$$
Using  the KLT relationship with two such tree amplitudes,
$A_4^{\rm tree}(\eps_1,\eps_2,\eps_3,\eps_4)$ and
$A^{\rm tree}_4(\bar{\eps}_1,\bar{\eps}_2,\bar{\eps}_3,\bar{\eps}_4)$,
we form the  combination
$$
M^{\rm tree, P}_4(
\eps_1,\bar{\eps}_1;
\eps_2,\bar{\eps}_2;
\eps_3,\bar{\eps}_3;
\eps_4,\bar{\eps}_4) =
 -\frac{i s_{12}}{4} \, A^{\rm tree}_4(\eps_1,\eps_2,\eps_3,\eps_4) \,
A^{\rm tree}_4(\bar{\eps}_1,\bar{\eps}_2,\bar{\eps}_4,\bar{\eps}_3)
\equn
$$
which
we will refer to as a {\it primitive} amplitude.
This primitive amplitude corresponds to
the scattering of massless states described by polarisation tensors
$\eps^{ab}_i=\eps_i^a \bar{\eps}_i^{b}$,
which in general will {\it not} be irreducible states
but will be a
combination of the polarisation tensors of a graviton, a two-form and a
scalar
$$
\eps_i^{a}\bar{\eps}_i^{b}
= \left[
\half ( \eps_i^{a}\bar{\eps}_i^{b}+\eps_i^{b}\bar{\eps}_i^{a} )
-{\eta^{ab} \over D} \eps_i \cdot \bar{\eps}_i \right]
+\half
(\eps_i^{a}\bar{\eps}_i^{b}-\eps_i^{b}\bar{\eps}_i^{a})
+{\eta^{ab} \over D} \eps_i \cdot \bar{\eps}_i
\equn
$$
Consequently the scattering amplitudes of irreducible states such as the graviton will
be a linear combination of these primitive amplitudes.

We can also use the KLT for states without polarisation tensors,
i.e. scalars.
Specifically we can calculate the amplitude for two
scalars and two gravitons where the graviton polarisation tensors are
$$
\epsilon_2^{ab} =
{1 \over \sqrt{2} }\left( \eps_2^a \bar{\eps}_2^b
+\eps_2^b \bar{\eps}_2^a \right)
\hskip 1.0 truecm
\epsilon_3^{ab} =
{1 \over \sqrt{2} }\left( \eps_3^a \bar{\eps}_3^b
+\eps_3^b \bar{\eps}_3^a \right)
$$
where (for simplicity) $\eps_i \cdot \bar{\eps}_i=0$,
from the primitive
amplitudes involving
two scalars  and two non-trivial polarisations
$$
\eqalign{ M^{\rm tree}_4(1_s ;  2_g ; 3_g ; 4_s ) =&
\half \bigg[ M^{\rm tree,P }_4(s_1; \eps_2,\bar{\eps}_2; \eps_3,\bar{\eps}_3; s_4)
+M^{\rm tree , P }_4(s_1; \eps_2,\bar{\eps}_2; \bar{\eps_3},\eps_3; s_4)
\cr & + M^{\rm tree, P}_4( s_1; \bar{\eps}_2,\eps_2; \eps_3,\bar{\eps}_3; s_4)
+M^{\rm tree, P }_4( s_1; \bar{\eps}_2,\eps_2; \bar{\eps}_3,\eps_3; s_4) \bigg]
 \cr}
\equn
$$ 
where the primitive amplitudes may be calculated using the KLT
relations.  (Formalisms where one need not symmetrise between left and right helicities also exist for gravity~\cite{BernGrant}.) 
Of the four terms in this expression there is a doubling up
to give two separate terms
because of a total symmetry between left and right, 
$$M^{\rm tree, P}_4(
\eps_1,\bar{\eps}_1;
\eps_2,\bar{\eps}_2;
\eps_3,\bar{\eps}_3;
\eps_4,\bar{\eps}_4)
=M^{\rm tree, P}_4(
\bar{\eps}_1,\eps_1;
\bar{\eps}_2,\eps_2;
\bar{\eps}_3,\eps_3;
\bar{\eps}_4,\eps_4)
\equn
$$

Primitive amplitudes can also generate
amplitudes with external two-forms by
antisymmetrising.
For example for the graviton scattering we have
$$
M^{\rm tree}_4(1_s; 2_g ; 3_g ; 4_s ) =
M^{\rm tree, P}_4( s_1;
\eps_2,\bar{\eps}_2; \eps_3,\bar{\eps}_3; s_4)
+M^{\rm tree, P}_4(
s_1; \eps_2,\bar{\eps}_2; \eps_3,\bar{\eps}_3;
s_4)
\equn
$$
while for the antisymmetric tensor
$$
M^{\rm tree}_4(1_s; 2_B ; 3_B ; 4_s ) =
M^{\rm tree, P}_4( s_1;
\eps_2,\bar{\eps}_2; \eps_3,\bar{\eps}_3; s_4)
-M^{\rm tree, P}_4( s_1; \eps_2,\bar{\eps}_2; \eps_3,\bar{\eps}_3; s_4)
\equn
$$
As we can see,
encompassed in the primitive amplitudes are the
contributions corresponding to a variety of Feynman diagrams.
Symmetrising or antisymmetrising projects to two rather different subsets of these. Diagrammatically

\SetScale{1}

\begin{center}
\begin{picture}(440,60)(0,0)
\Text(0,30)[l]{$M^{\rm tree, P}_4(s;\eps_2,\bar{\eps}_2;\eps_3,\bar{\eps}_3,s)
+ M^{\rm tree, P}_4(s;\eps_2,\bar{\eps}_2;\bar{\eps}_3,\eps_3;s) =$}
\SetColor{Red}
\Gluon(260,15)(275,30){2}{4}
\Gluon(260,45)(275,30){2}{4}
\Gluon(275,30)(305,30){2}{5}
\SetColor{Blue}
\Line(320,15)(305,30)
\Line(320,45)(305,30)
\SetColor{Red}
\Gluon(350,0)(365,15){2}{4}
\SetColor{Blue}
\Line(380,0)(365,15)
\Line(365,15)(365,45)
\SetColor{Red}
\Gluon(350,60)(365,45){2}{4}
\SetColor{Blue}
\Line(380,60)(365,45)
\SetColor{Red}
\Gluon(410,15)(425,30){2}{4}
\SetColor{Blue}
\Line(440,15)(425,30)
\SetColor{Red}
\Gluon(410,45)(425,30){2}{4}
\SetColor{Blue}
\Line(440,45)(425,30)
\Text(340,30)[]{$+$}
\Text(395,30)[]{$+$}
\Text(322,13)[lt]{$1_s$}
\Text(258,13)[rt]{$2_g$}
\Text(258,47)[rb]{$3_g$}
\Text(322,47)[lb]{$4_s$}
\Text(382,0)[lt]{$1_s$}
\Text(348,0)[rt]{$2_g$}
\Text(348,60)[rb]{$3_g$}
\Text(382,60)[lb]{$4_s$}
\Text(442,12)[lt]{$1_s$}
\Text(416,12)[rt]{$2_g$}
\Text(416,49)[rb]{$3_g$}
\Text(442,49)[lb]{$4_s$}
\end{picture}
\end{center}

\begin{center}
\begin{picture}(440,60)(0,0)
\Text(0,30)[l]{$M^{\rm tree, P}_4(s;\eps_2,\bar{\eps}_2;\eps_3,\bar{\eps}_3,s)
- M^{\rm tree, P}_4(s;\eps_2,\bar{\eps}_2;\bar{\eps}_3,\eps_3;s) =$}
\SetColor{Green}
\Photon(260,15)(275,30){2}{4}
\Photon(260,45)(275,30){2}{4}
\SetColor{Red}
\Gluon(275,30)(305,30){2}{5}
\SetColor{Blue}
\Line(320,15)(305,30)
\Line(320,45)(305,30)
\Text(322,13)[lt]{$1_s$}
\Text(258,13)[rt]{$2_B$}
\Text(258,47)[rb]{$3_B$}
\Text(322,47)[lb]{$4_s$}
\end{picture}
\end{center}


\section{One-Loop Amplitudes}

There are a variety of techniques for calculating on-shell
loop amplitudes,
often more efficient than a Feynman diagram approach.
In our calculations, we use two quite different
alternates to Feynman diagrams.

\subsection{Cutkosky Cutting Technique}

The optical theorem leads to the Cutkosky cutting rules~\cite{Cutting1}
in field
theory and it is possible to use these rules to determine amplitudes
{\it provided} one evaluates the cuts to ``all orders in
$\epsilon$''~\cite{Cutting2,BDKReview,DunbarTurner,LivingReview}.
(This is within the context of dimensional
regularisation where amplitudes are evaluated in
$D=2 N -2\epsilon$.)
These all-$\epsilon$ results allow a complete reconstruction of the
amplitude for a range of dimensions.

The cuts of a loop amplitude can be expressed in terms of
amplitudes containing fewer loops.  For example, the two-particle
cut of a one-loop
four-point amplitude in the $s$-channel,
as
shown in figure~\ref{CuttingFigure},
can be expressed as a product of tree amplitudes
$$
 -i \left. {\rm Disc}~M^{\rm 1-loop}_4(1 ,2 ,3 ,4)
\right|_{s-cut} \ =
 \int d LIPS\sum_{ {\rm internal}\atop {\rm states,s}}
\,\Mtree_4(-L_1^s,1 ,2 ,L_3^s)\,\Mtree_4(-L_3^s, 3 ,4 ,L_1^s )
\equn
$$
where the $dLIPS$ denotes integrating over the exchange momenta
$L_i$ subject to on-shell constraints
and where $L_3 = L_1 - k_1 - k_2$ and the sum runs over all states
crossing the cut.
The right-hand-side can be rewritten as the cut of a covariant integral
$$
\sum_{{\rm internal}\atop {\rm states,s}}
\left. \int {d^D L_1 \over (2 \pi)^D} \,
  {i\over L_1^2} \,
{ M}_{4}^\tree(-L_1^s,1,2,L_3^s)
\times
\,{i \over L_3^2}\,
{ M}_{4}^\tree (-L_3^s,3,4,L_1^s) \right|_{\rm s- cut}
\equn\label{BasicCutEquation}
$$
We label $D$-dimensional momenta with capital letters and
four-dimensional components with lower case letters.
We apply the on-shell
conditions, $L_1^2 = L_3^2 = 0$, to the amplitudes appearing in the cut
even though the loop momentum is unrestricted; only functions with a
cut in the given channel under consideration are determined in
this way.  By evaluating expressions with the correct cut in all
channels the full amplitude is determined.

\SetScale{1}
\begin{center}
\begin{picture}(300,120)(0,0)
\DashLine(150,15)(150,105){3}
\SetColor{Blue}
\ArrowLine(120,30)(120,90)
\ArrowLine(120,90)(147,97)
\ArrowLine(153,97)(180,90)
\ArrowLine(180,90)(180,30)
\ArrowLine(180,30)(153,23)
\ArrowLine(147,23)(120,30)
\SetColor{Red}
\Gluon(120,90)(95,115){4}{3}
\Gluon(120,30)(95,5){4}{3}
\Gluon(180,90)(205,115){4}{3}
\Gluon(180,30)(205,5){4}{3}
\Text(95,5)[rt]{$1^{}$}
\Text(95,115)[rb]{$2^{}$}
\Text(205,115)[lb]{$3^{}$}
\Text(205,5)[lt]{$4^{}$}
\SetColor{Black}
\LongArrowArcn(150,80)(30,120,60)
\LongArrowArcn(150,40)(30,300,240)
\Text(150,115)[cb]{$L_3$}
\Text(150,5)[ct]{$L_1$}
\end{picture}
\vskip 0.5 cm
{\rm Figure 4.1: The s-channel cut}
\label{CuttingFigure}
\end{center}

When evaluating graviton amplitudes in this way, the
the KLT expressions may be used to replace the
graviton tree amplitudes appearing in the cuts with products of gauge
theory amplitudes.
As an example, consider the specific case of a four graviton amplitude
where all four external (outgoing) states have the polarisation tensor
$\epsilon^{++}_{ab}$.  Consider the one-loop amplitude where a complex
scalar circulates in the loop.  This amplitude has non-zero cuts in all three
channels however, if we evaluate the $s$-channel the
others may be obtained by symmetry.

The tree amplitudes we need are for two gravitons and two scalars, and
these may be determined using the KLT relationships from gauge theory
partial
tree amplitudes with two external complex scalar legs and two gluons.  This
partial amplitude is
$$
\eqalign{
& A_4^\tree(-L_1^s,1^+,2^+,L_3^s) =  -i
{\mu^2\spb1.2\over\spa1.2 [( \ell_1 -k_1)^2 -\mu^2] }
\cr}
\equn$$
where we split the momenta into their four dimensional components and
$(D-4)$-dimensional components,
$L_1 = \ell_1 + \mu_1$.
Since the external momenta are purely four dimensional, $\mu_1=\mu_3\equiv\mu$.
The overall factor of $\mu^2$
appearing in these tree amplitudes indicates that they vanish in the
four-dimensional limit, in accord with a supersymmetry Ward
identity~\cite{SWI}.
Calculating the gravity amplitude,
$$
\eqalign{
 M_4^\tree(-L_1^s, 1_g^{++}, 2_g^{++}, L_3^s)
&= -\frac{is}{4} \, A_4^\tree(-L_1^s,1^+,2^+,L_3^s)  A_4^\tree(-L_1^s,2^+,1^+,L_3^s)
\cr
&={is\over 4} { (\mu^2)^2 \spb1.2^2 \over \spa1.2^2[( \ell_1 -k_1)^2 -\mu^2]
[( \ell_1 -k_2)^2 -\mu^2] }
\cr
&=-{ i\over 4}  \biggl({\mu^2\spb1.2\over\spa1.2}\biggr)^2 \times
 \biggl[{1\over (\ell_1 -k_1)^2-\mu^2}
 + {1\over (\ell_1 -k_2)^2-\mu^2}\biggr]
\cr}
\equn
$$
where we have used the fact that $s + [( \ell_1 -k_1)^2 -\mu^2]+[( \ell_1 -k_2)^2 -\mu^2]=0$.

Thus we have
$$
\eqalign{
{i\over L_1^2} \,
& { M}_{4}^\tree(-L_1,1_g^{++},2_g^{++},L_3)
 \times
\,{i \over L_3^2}\,
{ M}_{4}^\tree (-L_3,3_g^{++},4_g^{++},L_1)
\cr
=& {1 \over 16}{1\over L_1^2 L_3^2}
\, \biggl(
{  (\mu^2)^2 \spb1.2\spb3.4\over\spa1.2\spa3.4 }\biggr)^2
\biggl[{1\over (\ell_1 -k_1)^2-\mu^2}
 + {1\over (\ell_1 -k_2)^2-\mu^2}\biggr]
\biggl[{1\over (\ell_1 -k_3)^2-\mu^2}
 + {1\over (\ell_1 -k_4)^2-\mu^2}\biggr]
\cr
=&
{1 \over 16}\biggl(
{  \spb1.2\spb3.4\over\spa1.2\spa3.4 }\biggr)^2
{  (\mu^2)^4 \over L_1^2  L_3^2}
\biggl[{1\over (\ell_1 -k_1)^2-\mu^2}
 + {1\over (\ell_1 -k_2)^2-\mu^2}\biggr]
\biggl[{1\over (\ell_1 -k_3)^2-\mu^2}
 + {1\over (\ell_1 -k_4)^2-\mu^2}\biggr]
\cr}
\equn
$$
In this expression there is an overall factor which does not depend
upon the loop momentum, this multiplies an expression which is the
product of four propagators with a factor of $(\mu^2)^4$ in the numerator.
The four terms  corresponds to the
four different orderings  of the legs $1234$ which have a $s$-cut.
This means
$$
\eqalign{
\int {d^D L_1 \over (2 \pi)^D}
{i\over L_1^2} \,
 { M}_{4}^\tree(-L_1,1_g^{++},2_g^{++},L_3)
 \times
\,{i \over L_3^2}\,
{ M}_{4}^\tree (-L_3,3_g^{++},4_g^{++},L_1) \bigg|_{s-cut}
 \hskip 3cm &
\cr
={2i   \over 16  (4\pi)^{D/2}} \biggl(
{  \spb1.2\spb3.4\over\spa1.2\spa3.4 }\biggr)^2
\times \left( I^D_{1234}[(\mu^2)^4]+ I^D_{1243}[(\mu^2)^4]
\right)\bigg|_{s-cut} &
\cr}
\equn
$$
where
$$
I^{D}_{1234}[X]
\equiv
-i
(4\pi)^{D/2}\int { d^{D} L  \over (2\pi)^D  }
{X \over L^2(L-k_1)^2(L-k_1-k_2)^2(L-k_1-k_2-k_3)^2}
\equn$$
and where the terms have doubled up since
$I^D_{1234}[(\mu^2)^4]=I^D_{2143}[(\mu^2)^4]$. This expression, by
construction has the correct $s$-cut.
The $t$ and $u$
channel cuts, in this case, can be obtained by relabeling
and a combined expression can be formed by noting
$$
{  \spb1.2\spb3.4\over\spa1.2\spa3.4 }
={  \spb1.3\spb2.4\over\spa1.3\spa2.4 }
={  \spb1.4\spb2.3\over\spa1.4\spa2.3 }
= {-st \over \spa1.2 \spa2.3 \spa 3.4 \spa4.1}
\equn
$$
which leads us to an
expression which has the correct cuts, to all orders in $\epsilon$,
$$
\eqalign{
\Mloop(1_g^{++},2_g^{++},3_g^{++},4_g^{++})
=
{ 2 i \over  16(4\pi)^{D/2}} &
\left( { st \over \spa1.2 \spa2.3 \spa 3.4 \spa4.1 }
\right)^2
\cr
& \times
\left(
I^D_{1234}[(\mu^2)^4]
+I^D_{1324}[(\mu^2)^4]
+I^D_{1243}[(\mu^2)^4]
\right)
\cr}
\equn$$
This integral $I^D_{1234}[(\mu^2)^4]$
can be converted to a ``shifted box integral''~\cite{DimShift}
$$
\eqalign{
I^{D}_{1234}[(\mu^2)^4]  &= { (D-4)(D-2)(D)(D+2) \over 16} I^{D+8}_{1234}
\cr}
\equn$$
This form of the amplitude is valid for all dimensions $D \geq 4$.
In even dimensions, for example $D=6,8,10,12$,
the shifted box integral is ultra-violet infinite,
$$
\eqalign{
\left. I^{D=4}_{1234}[(\mu^2)^4] \right|_{1/\epsilon} =&
{ -2\eps.2.4.6 \over 16 } \times { 1 \over 2  \eps}
{ 2s^2+st +2t^2  \over 2520 }
\cr
\left. I^{D=6}_{1234}[(\mu^2)^4] \right|_{1/\epsilon} =&
{ 2.4.6.8 \over 16} \times { 1 \over 6 \eps}
{ -u(3s^2-2st+3t^2) \over 30240}
\cr
\left. I^{D=8}_{1234}[(\mu^2)^4] \right|_{1/\epsilon}=&
{ 4.6.8.10 \over 16} \times { 1 \over 24 \eps}
{ 12s^4+3s^3t+2s^2t^2+3st^3+12t^4  \over
831600}
\cr
\left. I^{D=10}_{1234}[(\mu^2)^4] \right|_{1/\epsilon} =&
{ 6.8.10.12 \over 16} \times { 1 \over 120 \eps}
{
-u(10s^4-8s^3t+9s^2t^2-8st^3+10t^4)
\over 4324320}
\cr
\left. I^{D=12}_{1234}[(\mu^2)^4] \right|_{1/\epsilon} =&
{ 8.10.12.14 \over 16} \times { 1 \over 720 \eps}
{  (60s^6 +10s^5t +4s^4t^2+3s^3t^3+4s^2t^4+10st^5+60t^6)
\over 108972864000 }
\cr}
\equn
$$
which produce infinities in the amplitude $\CMloop(1_g^{++},2_g^{++},3_g^{++},4_g^{++})$
$$
\eqalign{
D=6: \ &
{ i \kappa^4 \over 8\eps (4\pi)^3  }
\left( { st \over \spa1.2 \spa2.3 \spa 3.4 \spa4.1 }
\right)^2
\left(  {s t u \over 504   }\right)
\cr
D=8: \ &
{ i \kappa^4 \over 8\eps(4\pi)^4  }
\left( { st \over \spa1.2 \spa2.3 \spa 3.4 \spa4.1 }
\right)^2
\left(  {(s^2 +t^2 +u^2)^2  \over 15120   }\right)
\cr
D=10: \ &
{ i \kappa^4 \over 8\eps (4\pi)^5  }
\left( { st \over \spa1.2 \spa2.3 \spa 3.4 \spa4.1 }
\right)^2
\left( { stu(t^2+u^2+s^2) \over 31680 }\right)
\cr
D=12: \ &
{ i \kappa^4 \over 8\eps(4\pi)^6  }
\left( { st \over \spa1.2 \spa2.3 \spa 3.4 \spa4.1 }
\right)^2
{ 1 \over 4612608} \left(
(s^2+t^2+u^2)^3 + { 109.8(stu)^2 \over 75}
\right)
\cr}
\equn$$
This amplitude is finite in $D=4$ but in all even dimension $D>4$ it
has non-vanishing ultra-violet infinities indication that the
subtraction of ultra-violet infinities will require the introduction of
counterterms in all even dimension larger than four thus precluding any possibility of a ``magic'' dimension where all infinities cancel.

Cutting techniques can also be used to provide exact expressions for
the amplitudes involving two-forms. In Appendix B we demonstrate the
computations leading to the expressions,
$$
\eqalign{
\CMloop(1_g^{IJ}, 2_g^{IK}, 3_g^{KL}, 4_g^{JL} )
=& \
+ { i\kappa^4 \over 16 (4\pi)^{D/2} } \bigg\{ \Bigl(
I^{D+8}_{1234}(s,t)+I^{D+8}_{1243}(s,u)
+I^{D+8}_{1423}(t,u)
\Bigr)
\cr
& \hskip 2.5cm
- \Bigl(  I_3^{D+6}(s)+ I_3^{D+6}(t)\Bigr)
+{1 \over 8} \left(I_2^{D+4}(s)+I_2^{D+4}(t) \right) \bigg\}
\cr
\CMloop(1_g^{IJ}, 2_g^{IK}, 3_B^{LJ}, 4_B^{LK} )
=& \
-{ i\kappa^4 \over 16 (4\pi)^{D/2} } \bigg\{
{1 \over 2}   I_3^{D+6}(s)
-{1 \over 8} I_2^{D+4}(s) \bigg\}
\cr
\CMloop(1_B^{IJ},2_B^{IK},3_B^{LJ},4_B^{LK} )
=& \
+ { i\kappa^4 \over 16 (4\pi)^{D/2} }\bigg\{
{1 \over 8} \left(
I_2^{D+4}(s)+I_2^{D+4}(u)
\right)\bigg\}
\cr}
\equn
$$
where $I_3^{D}(s)$ and $I_2^{D}(s)$ denote triangle and bubble integrals.

\subsection{String Based Rules}

The Bern-Kosower rules for evaluating QCD amplitudes~\cite{Long} arose
from the low-energy limit of string theory amplitudes.  In
conventional field theory they have been shown to be related to mixed
gauge choices~\cite{BernDunbar} and also to the ``World-line
formalism''~\cite{WorldLine}.
 The derivation of these rules and details of their validity and
application will not be repeated here since several reviews are
available~\cite{Bern:1992ad,BDKReview} .

Since String theory exists
most naturally in $D=10$ or $D=26$,
the rules may be trivially adapted
to $D\leq 10$, although the World-line formalism would suggest they are valid for all dimensions $D$.

The initial step in the rules is to draw all labeled $\phi^3$ diagrams,
excluding tadpoles.
The contribution from each labeled
$n$-point $\phi^3$-like diagram with $n_{\ell}$ legs attached to the
loop is
$$
\eqalign{
{\cal D} =
i { (-\kappa)^n \over (4\pi)^{D/2} }
 \Gamma(n_\ell-D/2)
&\int_0^1 dx_{i_{n_\ell-1}} \int_0^{x_{i_{n_\ell-1}}} dx_{i_{n_\ell-2}} \cdots
\int_0^{x_{i_3}}  dx_{i_2} \int_0^{x_{i_2}} dx_{i_1} \cr
& \times
{K^{}_{\rm red}(x_{i_1},\dots,x_{i_{n_\ell}}) \over
\Bigl(\sum_{l<m}^{n_\ell} P_{i_l}\cdot P_{i_m} \x {i_m}{i_l}
(1-\x {i_m}{i_l})\Bigr)^{n_\ell -D/2}}  \cr}
\equn
$$
where the ordering of the loop parameter integrals corresponds to the
ordering of the $n_\ell$ lines attached to the loop, $x_{ij} \equiv
x_i - x_j$. The $x_{i_m}$ are related to ordinary Feynman parameters
by $x_{i_m} = \sum_{j=1}^m a_j$.  This expression corresponds to the
expression one obtains in a Feynman diagram calculation {\it after}
evaluating the vertex algebra and carrying out the loop momentum
integral.  The string based rules are algebraic rules for determining
$K_{\rm red}$ - the ``reduced kinematic expression'', diagram by diagram
from an overall
kinematic expression
$$
\eqalign{
\hskip 1.5cm {\cal K} &=
\int \prod_{i=1}^n dx_i d \bar x_i \prod_{i<j}^n
\exp\biggl[ k_i\cdot k_j G_B^{ij} \biggr]
\exp \biggl[ (k_i\cdot\pol_j - k_j\cdot\pol_i) \, \Gbd^{ij}
         - \pol_i\cdot\pol_j\, \Gbdd^{ij} \biggr] \cr
& \hskip 1 cm \times
\exp \biggl[ (k_i\cdot\bar\pol_j - k_j\cdot\bar\pol_i) \, \Gbdb^{ij}
         - \bar\pol_i\cdot\bar\pol_j\, \Gbddb^{ij} \biggr]
\exp \biggl[- ( \pol_i\cdot\bar\pol_j + \pol_j\cdot\bar\pol_i )
\,  H_B^{ij} \biggr]
\biggr|_{\rm multi-linear}
\cr}
\equn
$$
where the `multi-linear' indicates that only the terms linear in all
$\pol_i$ and $\bar\pol_i$ are included.  The graviton polarization
tensor is reconstructed from the
$\pol_i^a \bar\pol_i^b$ as before.
Although the above expression contains much information in string
theory, when one takes the
infinite string tension limit~\cite{Long,BDKReview} it
should
merely be regarded as a function
which contains all the information
necessary to generate $K_{\rm red}$ for all graphs.
The utility of the string based method
partially lies in this compact
representation (which is valid for arbitrary numbers of legs).
The existence of an
overall function which reduces to the Feynman parameter polynomial for each
diagram is one of the most useful features of the string based rules.

As an example of the string based technique we can look at the
four-point amplitude
\break $\CMloop(1_g^{--},2_g^{++},3_g^{++},4_g^{++})$
with a complex scalar circulating in the loop.
This choice of helicity simplifies the kinematic expression considerably
and we can deduce that the amplitude is given by
(This was first
calculated using the string-based technique of Bern and
Kosower~\cite{Long,BernDunbar} applied to quantum gravity calculations
\cite{BDS,DunNor})
$$
\CMloop(1_g^{--},2_g^{++},3_g^{++},4_g^{++}) =
 {2 i \kappa^4 \over16}
\left({{\spb2.4}^2st^2\over\spb1.2\spa2.3\spa3.4\spb4.1}\right)^2
\left(
{\cal I}_s^{(a)}
+{\cal I}_s^{(b)}
+{\cal I}_s^{(c)}
+{\cal I}_s^{(d)}
+{\cal I}_s^{(e)}
\right)
\equn$$
where
$$
\eqalign{
{\cal I}_s^{(a)} &=
{\Gamma(4-D/2)\over (4\pi)^{D/2}}
\int_0^1 \prod_{i=1}^4 da_i \delta\left(1-\sum_{i=1}^4
a_i\right)
{\left((a_1+a_2)a_3^2a_4\right)^2\over(-sa_1a_3-ta_2a_4)^{4-D/2}}
\cr
{\cal I}_s^{(b)} &=
{\Gamma(4-D/2)\over (4\pi)^{D/2}}
\int_0^1 \prod_{i=1}^4 da_i \delta\left(1-\sum_{i=1}^4
a_i\right)
{\left((a_1+a_2)a_3^2a_4\right)^2
\over(-ua_1a_3-ta_2a_4)^{4-D/2}}
\cr
{\cal I}_s^{(c)} & =
{\Gamma(4-D/2)\over (4\pi)^{D/2}}
\int_0^1 \prod_{i=1}^4 da_i \delta\left(1-\sum_{i=1}^4
a_i\right)
{\left((a_1+a_2)(a_3+a_4)a_3^2a_4\right)^2
\over(-sa_1a_3-ua_2a_4)^{4-D/2} }
\cr
{\cal I}_s^{(d)} & =
-{\Gamma(3-D/2)\over (4\pi)^{D/2}}
{1 \over s}
\int_0^1 \prod_{i=2}^4 \,da_i \delta\left(1-\sum_{i=2}^4
a_i\right)
{\left(a_2a_3a_4\right)^2 \over(-sa_2a_3)^{3-D/2}}
\cr
{\cal I}_s^{(e)} & =
-{\Gamma(3-D/2)\over (4\pi)^{D/2}}
{1 \over u}
\int_0^1 \prod_{i=2}^4 \,da_i \delta\left(1-\sum_{i=2}^4
a_i\right)
{\left(a_2a_3a_4\right)^2 \over(-ua_2a_3)^{3-D/2}}
\cr}
\equn$$
These yield a finite result in $D=4$, however in higher dimensions they give
rise to ultra-violet infinities
$$
\eqalign{
D=6: \ & {i\kappa^4 \over 8\eps (4\pi)^3}
\left({{\spb2.4}^2s^2t\over\spb1.2\spa2.3\spa3.4\spb4.1}\right)^2
\times \left(  {t \over5040  s u }\right)
\cr
D=8: \ & {i\kappa^4\over 8\eps (4\pi)^4}
\left({{\spb2.4}^2st^2\over\spb1.2\spa2.3\spa3.4\spb4.1}\right)^2
\times \left( {0}\right)
\cr
D=10: \ & {i\kappa^4\over 8\eps (4\pi)^5 }
\left({{\spb2.4}^2st^2\over\spb1.2\spa2.3\spa3.4\spb4.1}\right)^2
\times  \left(
0
\right)
\cr
D=12: \ & {i\kappa^4\over 8\eps (4\pi)^6  }
\left({{\spb2.4}^2st^2\over\spb1.2\spa2.3\spa3.4\spb4.1}\right)^2
\times  \left(  { t^2 \over 37837800}  \right)
\cr}
\equn$$
Again we see the presence of ultra-violet
infinities in higher dimensional one-loop amplitudes however, in this
case they vanish for $D=8$ and $D=10$ indicating that the counterterms
must have a form which does not contribute to this amplitude.
In fact, as we see later, all of the possible counterterms which are
consistent with
the symmetries of gravity in $D=8$ have vanishing contributions for this particular helicity configuration.

Including the two examples we have just calculated,
there are sixty-nine independent, non-vanishing helicity
configurations for four external gravitons in $D> 4$ dimensions.
(Amplitudes not listed are either zero to all orders or obtainable from the list by relabeling or complex conjugation.) The
tree amplitudes of these are listed in Appendix~A.  The string based
rules may be used to calculate the loop amplitude for
any of these. The
ultra-violet infinities in $D=6,8,10$ for the first thirty-one of
these amplitudes is given in Appendix~C.  This subset of the
amplitudes provides more than sufficient information to determine the
counterterms necessary to cancel the infinities in four graviton amplitudes.
In the following sections we shall detail this process for $D=6,8$.

The string based rules can be applied to determine the contributions
to amplitudes
for particle types other
than that of scalars circulating in the loop. This corresponds to
applying different algebraic rules in determining $K_{\rm red}$.
This will allow us to
determine infinities in the amplitudes and hence
the counterterms induced by other particle types.

\section{Counterterms}

In this section we enumerate the possible independent counterterms in
six and eight dimensions and show how the results of the one-loop
amplitude calculations determine the various coefficients.

\subsection{Symmetries}

In general, graviton scattering amplitudes,
in $D$ dimensions at $L$ loops,
are rendered ultra-violet finite by the introduction
of counterterms of the
form
$$
\nabla^n R^m
\equn
$$
where $n+2m = (D-2)L+2$
and
we have suppressed the indices on $R$.
$R$ may stand either for
the Riemann tensor, $R_{abcd}$, the Ricci tensor $R_{ab}\equiv
g^{cd}R_{acbd}$ or the curvature scalar $R \equiv g^{ab}R_{ab}$.
Although, there are  a large number of tensor structures which may
appear, fortunately, the symmetries of the Riemann tensor reduce these
considerably. Firstly, there are the basic symmetries of $R_{abcd}$
$$
R_{abcd}=-R_{bacd}=-R_{abdc}=R_{cdab}
\equn
$$
and the cyclic symmetry,
$$
R_{abcd}+R_{acdb}+R_{adbc}=0
\equn
$$
Secondly, we have the Bianchi identity for $\nabla_e R_{abcd}$,
$$
\nabla_e R_{abcd}
+\nabla_c R_{abde}
+\nabla_d R_{abec}
=0
\equn
$$
There are also ``derivative symmetries'' which involve two covariant derivatives,
$$
\eqalign{
\nabla_{e}\nabla_{f} R_{abcd}
-\nabla_{f}\nabla_{e} R_{abcd}
&= R^g{}_{aef} \, R_{gbcd}
+ R^g{}_{bef} \, R_{agcd}
+ R^g{}_{cef} \, R_{abgd}
+ R^g{}_{def} \, R_{abcg}
\cr
\nabla^2 R_{abcd}
& =
 2 \, R^{f}_{\ a c e} \, R^{e}_{\ d b f}
- 2 \, R^{f}_{\ b c e} \, R^{e}_{\ d a f}
- R^{e}_{\ d a b} \, R_{c e}
+ R^{e}_{\ c a b} \, R_{d e}
\cr
& \hskip 1cm
+ \nabla_{c} \nabla_{a} R_{b d} - \nabla_{c} \nabla_{b} R_{a d}
- \nabla_{d} \nabla_{a} R_{b c} + \nabla_{d} \nabla_{b} R_{a c}
\cr}
\equn\label{DerivativeSymmetry}
$$

\noindent
These symmetries will be used to determine the minimal set of
inequivalent counterterms.

\subsection{Graviton and Two-Form Scattering in $D=6$}

By power counting the
possible counterterms in $D=6$ are of the form $R^3$ or $\nabla^2 R^2$.
The independent terms involving $R_{abcd}$, $R_{ab}$ and $R$
are \cite{R3D4,Fulling},
$$
\eqalign{
T_1&=\nabla_{a} R \nabla^{a} R
\cr
T_2&=\nabla_{a} R_{bc} \nabla^{a} R^{bc}
\cr
T_3&=\nabla_{e} R_{abcd} \nabla^{e} R^{abcd}
\cr
T_4&=\nabla_{c} R_{ab} \nabla^{b} R^{ac}
\cr
T_5&= R^3
\cr
T_6&= R R_{ab}  R^{ab}
\cr
T_7&= R R_{abcd}  R^{abcd}
\cr
T_8&= R_{abcd} R^{abce} R^{d}_{\ e}
\cr
T_9&=  R_{abcd} R^{ac} R^{bd}   
\cr
T_{10}&=R_{a}{}^b R_b{}^cR_c{}^a
\cr
T_{11} &= R^{ab}{}_{cd}R^{cd}{}_{ef}R^{ef}{}_{ab}  \;\;\;
\cr
T_{12}&=R_{abcd}R^{a}{}_e{}^c{}_f R^{bedf}
\cr}
\equn$$
(For $D=4$ only ten of this set are independent.)
For the case of pure gravity, the counterterm structure can be
represented as a single counterterm with a numerical coefficient. This
numerical coefficient has been
calculated previously~\cite{d6infinity}.  We
review the argument leading to the conclusion that a single
counterterm is sufficient.
When matter is
coupled to gravity this conclusion no longer follows.

For pure gravity the equation of motion is
$$
R_{ab}=0
\equn$$
Hence terms involving the Ricci tensor or curvature scalar
$$
R_{ab} \, X^{ab}   \;\;,  \;\; \;\;  R \, X = R_{ab} \, (g^{ab} X)
\equn$$
will not contribute to the $S$-matrix and such terms can be
discarded when calculating the counterterms. If calculating an off-shell
object, such counterterms can, and do, appear.
Ignoring such terms leaves us with three tensors - $T_3$, $T_{11}$ and $T_{12}$.
The term $T_3$
$$
\nabla_e R_{abcd}
\nabla^e R^{abcd} =-  R_{abcd} \nabla^2 R^{abcd}
\equn$$
can be rearranged using the  identity
in eq.~(\ref{DerivativeSymmetry})
into terms involving the Ricci tensor plus
cubic terms in the Riemann tensor.
Thus for pure gravity this term is equivalent to a combination of
$T_{11}$ and $T_{12}$ and can thus be
eliminated from the list of inequivalent counterterms.

In six dimensions the scalar topological density can be written
$$
\delta^{\ a \, b \, c d e f}_{[mnpqrs]} R^{mn}{}_{ab} R^{pq}{}_{cd}R^{rs}{}_{ef}
\equn$$
which implies the combination
$$
\sum_{i=5}^{12} a_i T_i \equiv 0
\equn$$
is topological for some coefficients $a_i$.
Hence for pure gravity amplitudes we can replace
$T_{12}$ for $T_{11}$ (or vice versa).  Thus we are led to the fact that
the counterterm can be taken as a single tensor with a coefficient.
This argument also applies to the two-loop case of pure gravity in
$D=4$~\cite{GoroffSagnotti}.  In both this calculation and that of
$D=6$ pure gravity the counterterm was chosen to be
$$
c \int d^6 x \, R^{ab}{}_{cd}R^{cd}{}_{ef}R^{ef}{}_{ab}
\equn$$

For our case we are considering gravity amplitudes with scalar loops. For gravity coupled to matter the field equation is
$$
R_{ab}-{ \kappa^2  \over 4 (D-2) }
\bigg[ g_{ab} T_c{}^{c} -(D-2) T_{ab} \bigg]
 =0
\equn\label{FieldEquationB}
$$
so counterterms involving the Ricci tensor can no longer trivially be
dropped.
However we shall always be able to make the replacement
$$
R_{ab} X^{ab} \longrightarrow { \kappa^2 \over 4 (D-2) } \bigg[ g_{ab}
T_c{}^{c} -(D-2) T_{ab} \bigg] X^{ab}
\equn
$$
without changing
the $S$-matrix. The right-hand-side  involves at least two matter fields
and thus does not contribute to pure graviton amplitudes, but may
contribute to amplitudes involving two gravitons and two matter
fields.  Thus we may still neglect counterterms involving the Ricci
tensor provided we are restricting attention to external gravitons.
(This is similar to the situation in $D=4$.)
Thus we are led to the same conclusion as for pure gravity in that
the infinities can be renormalised by a single counterterm.

Knowing the counterterm is unique we can
fix $c$ from a single amplitude - providing the amplitude is non-zero
for that term. Either of the amplitudes we presented earlier,
$\CMloop(1_g^{--},2_g^{++},3_g^{++},4_g^{++})$ and
$\CMloop(1_g^{++},2_g^{++},3_g^{++},4_g^{++})$,
would be sufficient to determine the coefficient.
Thus from either
of these amplitudes we can confirm the
non-vanishing of the counterterm and extract the coefficient
(The value we obtain matches that of {\it all} the amplitudes we calculate in Appendix C.)
$$
c = -{ 1  \over (4 \pi)^3 \epsilon } \times { 1 \over 15120} \
\equn$$
This counterterm will make amplitudes with a complex scalar loop and
external gravitons finite.
As we shall see later the pure gravity case
will simply be $9/2$ times this. Multiplying this by a factor of
$9/2$ does indeed give the previously calculated result.

When considering amplitudes other than pure graviton scattering the
single counterterm above will not be sufficient to cancel the
infinities.  To fully determine all the coefficients it would be
necessary to compute six-point amplitudes involving, for example, six
$B_{ab}$ or scalar fields since terms such as $T_5$ can be replaced by
tensors involving six matter fields. Alternatively one could say
these terms are unnecessary to cancel the infinities in four-point
amplitudes.
However, some of the counterterms involving the Ricci tensor
will need to be introduced to cancel infinities in four-point
amplitudes where some of the external states are matter states.
For example, if we consider
amplitudes  with two external
two-forms and two external gravitons (still with a scalar loop) .
This is computationally fairly straightforward since
within string
theory, the graviton and antisymmetric two-form are very closely tied
together. Using string based rules this means that the amplitudes
involving two-forms are very closely related to the amplitudes
involving gravitons - the amplitude is formed from the
same primitive amplitudes but
with different signs. From a
more traditional field theory view it would also be relatively
easy. The antisymmetric tensor does not couple to the scalar field
directly so the Feynman graphs are of the form,

\vskip 0.5 truecm
\SetScale{1}

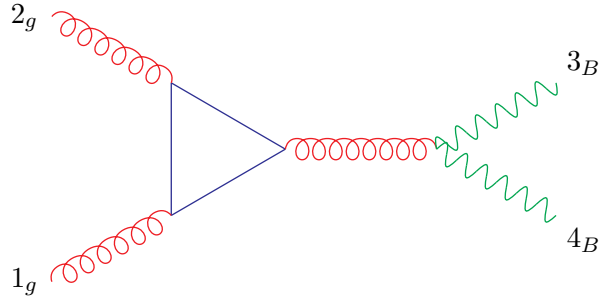
\begin{center}
\begin{picture}(300,100)(0,0)
\SetColor{Blue}
\Line(100,25)(100,75)
\Line(100,25)(143,50)
\Line(100,75)(143,50)
\SetColor{Red}
\Gluon(143,50)(200,50){4}{8}
\Gluon(55,0)(100,25){4}{6}
\Gluon(55,100)(100,75){4}{6}
\SetColor{Green}
\Photon(245,25)(200,50){4}{6}
\Photon(245,75)(200,50){4}{6}
\Text(50,0)[r]{$1_g$}
\Text(50,100)[r]{$2_g$}
\Text(250,80)[lb]{$3_B$}
\Text(250,20)[lt]{$4_B$}
\end{picture}
\end{center}

\vskip -0.25 truecm
\noindent
As we can see from this diagram, this is equivalent
to  probing the off-shell graviton three-point function.
In six
dimensions if we trace the equations of motion eq.~(\ref{FieldEquationB})
$$
\eqalign{
g^{ab} R_{ab} &={ \kappa^2 \over 4 (D-2) }
g^{ab} \bigg[ g_{ab} T_c{}^c -(D-2) T_{ab} \bigg]
\cr
&= g^{ab} \left( \frac{\kappa}{2} \right)^2 \, \left[
\frac{2}{3 (D-2)} \, g_{ab} \, F_{cde} \, F^{cde}
- F_{a cd} \, F_{b}^{\ cd}
- 2 \, \partial_{a} \phi^* \, \partial_{b} \phi \right]
\cr
&= \left( \frac{\kappa}{2} \right)^2   \left( - 2 \, \partial_{a} \phi^* \, \partial^{a} \phi \right)
\cr}\equn
$$
the two-form is eliminated.  Hence a counterterm involving the
curvature scalar $R$ can be replaced by a counterterm quadratic in the
scalar field and such counterterms will not contribute to amplitudes
with external two-forms. The counterterms which will give
non-vanishing contributions to amplitudes with two gravitons and two
2-forms are  $T_3$, $T_{8}$, $T_{11}$ and $T_{12}$.  This set of
four tensors are not independent and we can use the previous argument
for eliminating $T_3$ and $T_{12}$ from the minimal set of tensors
leaving the two counterterms $T_8$ and $T_{11}$.  The coefficient of
$T_{11}$ is fixed by the amplitudes with four external gravitons.
The counterterm
$$
T_8= R_{abcd} R^{abce} R^{d}_{\ e}
\equn
\label{hBcounterterms}
$$
is equivalent, for these amplitudes, to the tensor
$$
\left( \frac{\kappa}{2} \right)^2 \, R_{abc}^{\ \ \ d} \, R^{abce}
\, \left[ \frac{1}{6} \, g_{de} \, F_{fgh} \, F^{fgh}
- F_{dfg} \, F_{e}^{\ fg} \right]
\equn
$$
The infinities in a sufficiently large set of 
two graviton and two 2-form amplitudes are given
in the Appendix~D.  
Both  $T_8$ and $T_{11}$ contribute to these amplitudes and  
the following combination of these counterterms 
is needed to cancel both these infinities and those
of the four graviton amplitudes 
$$
-\frac{2}{(4 \pi)^3 \epsilon} \, \left\{ \frac{1}{30240} \, R^{ab}_{\ \ cd} \, R^{cd}_{\ \ ef} \, R^{ef}_{\ \ ab}
+ \frac{1}{1008} \left( \frac{\kappa}{2} \right)^2 \, R_{abc}^{\ \ \ d} \, R^{abce}
\, \left[ \frac{1}{6} \, g_{de} \, F_{fgh} \, F^{fgh}
- F_{dfg} \, F_{e}^{\ fg} \right] \right\}
\equn
$$
or if we wish to expressly write this as $R^3$ terms,
$$
-\frac{2}{(4 \pi)^3 \epsilon} \, \left\{ \frac{1}{30240} \, R^{ab}_{\ \ cd} \, R^{cd}_{\ \ ef} \, R^{ef}_{\ \ ab}
+ \frac{1}{1008} \, R_{abcd} \, R^{abce}R^{d}_{\ e}
\, \
\right\}
\equn
$$

We can further probe the counterterm action by calculating amplitudes
with four external two-forms.  Only the  tensors 
$$
T_1 =\nabla_{a} R \, \nabla^{a} R 
\hskip 2cm 
T_2 = \nabla_{a} R_{bc} \, \nabla^{a} R^{bc}
\equn
$$
will contribute to the scattering of four matter states.
Due to the fact that in $D=6$ the equation of motion 
for $R$ does not depend upon the two-form,  only $T_2$
contributes to that of four two-form states. 
Replacing $R_{ab}$ using the equation of motion, $T_2$ is
equivalent to
$$
\eqalign{
\hskip 2cm & \left( \frac{\kappa}{2} \right)^4 \, \nabla_a \, \left[ \frac{1}{6} \, g_{bc} \, F_{def} \, F^{def}
- F_{bde} \, F_{c}^{\ de} \right] \
\nabla^a \, \left[ \frac{1}{6} \, g^{bc} \, F_{ghi} \, F^{ghi}
- F^b_{\ gh} \, F^{cgh} \right]
\cr
&=\left( \frac{\kappa}{2} \right)^4 \, \left[ \partial_a \, \left( F_{bde} \, F_c^{\ de} \right)
\, \partial^a \, \left( F^b_{\ gh} \, F^{cgh} \right)
- \frac{1}{6} \, \partial_a \, \left( F_{bde} \, F^{bde} \right) 
\, \partial^a \, \left( F_{fgh} \, F^{fgh} \right) \right]
\cr}
\equn
$$
In Appendix~D the infinities for a set of four $2$-forms amplitudes are
presented. 
(The tensors $T_8$ and $T_{11}$ do not contribute to these amplitudes.)
Canceling these divergences fixes the coefficient of $T_2$,
$$
-\frac{2}{(4 \pi)^3 \epsilon} \, \frac{1}{1680} \,
\left( \frac{\kappa}{2} \right)^4 \, \left[ \partial_a \, \left( F_{bde} \, F_c^{\ de} \right)
\, \partial^a \, \left( F^b_{\ gh} \, F^{cgh} \right)
- \frac{1}{6} \, \partial_a \, \left( F_{bde} \, F^{bde} \right)
\, \partial^a \, \left( F_{fgh} \, F^{fgh} \right) \right]
\equn
$$
or equivalently
$$
-\frac{2}{(4 \pi)^3 \epsilon} \, \frac{1}{1680} \, \nabla_{a} R_{bc} \, \nabla^{a} R^{bc}
\equn
$$
Thus we can conclude that the counterterms necessary to make amplitudes with 
external gravitons or two-forms finite is
$$
-\frac{2}{(4 \pi)^3 \epsilon} \, \left\{ 
\frac{1}{30240} \, R^{ab}_{\ \ cd} \, R^{cd}_{\ \ ef} \, R^{ef}_{\ \ ab}
+ \frac{1}{1008} \, R_{abcd} \, R^{abce}R^{d}_{\ e}
+ \frac{1}{1680} \, \nabla_{a} R_{bc} \, \nabla^{a} R^{bc}
\, \
\right\}
\equn
$$
This is the counterterm generated by a complex scalar loop. In the
following section we examine the effects of having more complicated
particle types circulating in the loop.

\subsection{Counterterms Generated by More General Matter in $D=6$}

We have considered the counterterms generated by a single 
complex scalar.  The
results for more general matter combinations 
are very closely related to
the complex scalar case. If we have minimally coupled matter
the resultant counterterm is
$$
{\cal L } = {(N_B-N_F)\over 2 } \times  {\cal
 L }_{Scalar}
\equn$$
where $N_B$ is the number of bosonic degrees of freedom and $N_F$ is
the number of fermionic degrees of freedom. 

Our argument leading to this simple result is actually rather
complicated and uses the string based rules for graviton scattering.
These algebraic rules are for generating Feynman parameter integrals
as discussed in section~4.2. The rules can generate the contributions
for different matter combinations.

These rules are based upon string theory amplitudes 
in $D=10$. In $D=10$ language there are three underlying
types of contributions which we label  $[S]$, $[V]$ and $[F]$. 
In terms of particle content these correspond to 
the contributions from the
$1$, $8_v$ and $8_{s/c}$ representations of $SO(8)$ . 
For a closed string, which has left and right moving quanta, we have
the option of different $SO(8)$ representations for left and right.
So the rules for gravity generate contributions corresponding
to the product of 
these representations. 
In terms of particle content in $D=10$, this
corresponds to \def\ul#1{\underline{#1}}
$$
\eqalign{
[S;S] &=\ul{1}\otimes\ul{1} \equiv  \phi
\cr
[V;S] &=\ul{1}\otimes\ul{8}_v \equiv  A_a
\cr
[V;V] &=\ul{8}_v\otimes\ul{8}_v = \ul{1}+\ul{28}+\ul{35} 
\equiv \phi +A_{ab} +g_{ab}
 \cr
[S;F] & = \ul{1} \otimes\ul{8}_s = \ul{8}_s\equiv \lambda
 \cr
[V;F] & = \ul{8}_v\otimes\ul{8}_s= \ul{8}_c+\ul{56}_s \equiv
\psi_{a} + \lambda
\cr
[F;F] & = \ul{8}_s\otimes\ul{8}_s = \ul{1} +\ul{28}+\ul{35} \equiv
\phi +A_{ab} +A_{abcd}^{SD}
\cr
[F;\bar{F} ] &  = \ul{8}_s\otimes\ul{8}_c =\ul{8}_v +\ul{56}_v 
\equiv A_{a} +A_{abc}
\cr}
\equn$$
where $A_{abcd}^{SD}$ is a self-dual four-form field
and $\phi$ is a real scalar field.

In $D=10$ ``knowing'' the contributions from the above combinations of
matter does not actually allow us to determine the contribution due to
a single particle type - the five contributions $[S;S]$,$[V;S]$,
$[V;V]$, $[F;F]$ and $[F;\bar{F} ]$ cannot be disentangled to determine
the contributions from the six individual particles - $\phi$,
$g_{ab}$, $A_{a}$, $A_{ab}$, $A_{abc}$ and
$A_{abcd}^{SD}$.  However, in $D<10$ the five
basic combinations may be sufficient to determine the contributions
from all the particle types.  For example, in $D=4$ the antisymmetric
tensors will all reduce to combinations of three basic particles -
$\phi$, $A_{a}$ and $g_{ab}$ and in this case there is enough
information to (over)determine the three basic particle types.

In $D=6$ it transpires there is just enough information to determine
the contributions from the five basic bosonic particle types.
The string contributions,
for the bosonic terms, will be
$$
\eqalign{ [S;S] &=\ul{1}\otimes\ul{1} \equiv \phi \cr [
V;S]
&=\ul{1}\otimes\ul{8}_v \equiv A_{a} +\sum_{i=1}^4 \phi^i \cr 
[V;V]
&=\ul{8}_v\otimes\ul{8}_v \equiv \sum_{i=1}^{17} \phi^i +A_{ab}
+g_{ab} +\sum_{i=1}^8 A_{a}^i \cr 
[F;F] & =
\ul{8}_s\otimes\ul{8}_s \equiv \sum_{i=1}^{2} \phi^i +\sum_{i=1}^4
A^i_{a} +\sum_{i=1}^6 A^i_{ab} \cr 
[F;\bar{F} ] & =
\ul{8}_s\otimes\ul{8}_c = A_{abc}+ \sum_{i=1}^6 A^i_{ab}
+\sum_{i=1}^7 A^i_{a} +\sum_{i=1}^8 \phi^i \cr} 
\equn
$$ 
In this expression the combinations $[F;F]$ and $[F;\bar{F} ]$
correspond to identical sets of fields in $D=6$. This is because the two
type~II supergravities are  dual when compactified to $D < 10$
~\cite{Dai:ua}.
This then leaves us with four independent pieces of information.
This system is solvable because the three-form $A_{abc}$ is
dual to a vector $A_{a}$ in $D=6$ which means we have only
four independent field types - $\phi$, $A_{a}$ , $A_{ab}$ and $g_{ab}$. The fermionic terms easily allow us to determine the
contribution from the spinor $\lambda$ and gravitino $\psi_{a}$.

This tells us that if we can determine the string based contributions
we can solve to obtain the individual particle types.
The explicit results of calculations can be summarised for the ultra-violet infinities,
$$
\eqalign{
[V;S]=& +8 \, [S;S]
\cr
[V;S]=& +64 \, [S;S]
\cr
[F;F]=[F;\bar{F} ]=& +64 \, [S;S]
\cr
[S;F]=& -8 \, [S;S]
\cr
[V;F]=& -64 \, [S;S]
\cr}
\equn$$
The solution to this system of equations is that the infinities
in the one-loop amplitude from
a set of particles, $P$, is given by
$$
\left. M^{\rm 1-loop}_P \right|_{1/\eps}
={ (N_B-N_F)\over 2} \left. M^{\rm 1-loop}_{scalar} \right|_{1/\eps}
\equn$$
This relationship then obviously extends to the counterterm Lagrangian.

In four dimensions a similar string based argument holds, however, a
more elegant supersymmetry argument can be used to achieve equivalent
results.  In four dimensions the helicity amplitude with all-plus
helicities can be shown to vanish in any supersymmetric theory
$$
M^{\rm 1-loop,susy\; multiplet} (1^{++},2^{++},3^{++},4^{++})=0
\equn$$
This applies to all supersymmetries $N\geq1$. Since $N=1$ multiplets
are actually rather simple this relationship easily allows one to deduce
for all particle types
$$
M^{\rm 1-loop}_{P} (1^{++},2^{++},3^{++},4^{++})=
{ (N_B-N_F)\over 2} M^{\rm 1-loop}_{scalar}(1^{++},2^{++},3^{++},4^{++})
\equn$$
This relationship
is true of entire amplitudes and not merely the infinities.

\subsection{Graviton Scattering in $D=8$}

For $D=8$ the possible counterterms are of the form
$\nabla^4 R^2$,  $\nabla^2 R^3$ and $R^4$.
As we shall see, for external graviton amplitudes the set of
inequivalent counterterms can be
constructed entirely using the $R^4$ counterterms.

First, recall that
in purely graviton amplitudes terms involving
the Ricci tenser and curvature scalar do not contribute
leaving us with terms involving the Riemann tensor only.
Consider the terms quadratic in the Riemann tensor.
There are various possibilities for the indices of these tensors
but we can organise these into three types depending on
how many contractions the Riemann tensors have with each other.
Representatives of the three types are,
$$
\eqalign{
\nabla_e \nabla_f R_{abcd}
\nabla^c \nabla^d R^{abef}
\cr
\nabla_e \nabla_f R_{abcd}
\nabla^d \nabla^f R^{abce}
\cr
\nabla_e \nabla_f R_{abcd}
\nabla^e \nabla^f R^{abcd}
\cr
}
\equn$$
We have
chosen the representatives such that there are no contractions between the
derivatives and the tensor they act upon.  For such terms
we can use the Bianchi identity
$$
\nabla^a R_{abcd}=-  \nabla_c R_{abd}{}^a- \nabla_d R_{ab}{}^a{}_c
=  \nabla_c R_{bd} - \nabla_d R_{bc}
\equn$$
to equate this to Ricci tensors which we discard.
The order
of derivatives can also be changed - but at the expense of
$\nabla^2 R^3$ terms eq.~(\ref{DerivativeSymmetry}).
Thus the generic terms are equivalent to the
representative terms given.
We now show that these can be eliminated from the list of counterterms
in favour of $\nabla^2 R^3$ and $R^4$ terms.
Using the
antisymmetry of the $ef$ indices
the first term can be rewritten
$$
\nabla_e \nabla_f R_{abcd}
\nabla^c \nabla^d R^{abef}
=
{1 \over 2}
(\nabla_e \nabla_f R_{abcd}-\nabla_f \nabla_e R_{abcd} )
\nabla^c \nabla^d R^{abef}
\equiv  \nabla^2 R^3 \; {\rm terms }
\equn$$
For terms of second and  third type  we can
commute $\nabla$ (at the cost of creating
$\nabla^2 R^3$ terms) and integrate by parts to
bring the contracted derivatives together.
Acting on a Riemann tensor, equation~(\ref{DerivativeSymmetry})
shows that such terms are
equivalent to $\nabla^2 R^3$.

Turning to the $\nabla^2 R^3$ terms, there are four tensors involving
the Riemann tensor. The normal form of these is~\cite{Fulling},
$$
\eqalign{
S_1 =&
R_{abcd} \nabla_e R_f{}^a{}_g{}^c \nabla^e R^{fbgd}
\cr
S_2 =&
R_{abcd} \nabla^c R_{efg}{}^a \nabla^d R^{efgb}
\cr
S_3 =&
R_{abcd} \nabla^b R_{efg}{}^a \nabla^d R^{efgc}
\cr
S_4 =&
R_{abcd} R^a{}_{efg} \nabla^g \nabla^d R^{becf}
\cr}
\equn
$$
In manipulating these terms, commuting derivatives will
produce terms involving the Ricci or $R^4$ so this can be done at will.
Taking the $S_1$ first, integrating by parts yields
$$
S_1 =
R_{abcd} \nabla_e R_f{}^a{}_g{}^c \nabla^e R^{fbgd}
= - \nabla_e R_{abcd} R_f{}^a{}_g{}^c \nabla^e R^{fbgd}
- R_{abcd} R_f{}^a{}_g{}^c \nabla^2 R^{fbgd}
\equn
$$
so that
$$
S_1 = -\half \, R_{abcd} R_f{}^a{}_g{}^c \nabla^2 R^{fbgd}
\equn
$$
which is equivalent to $R^4$ terms.
The second term
$$
S_2 = R_{abcd} \nabla^c R_{efg}{}^a \nabla^d R^{efgb}
\equiv
-R_{abcd} R_{efg}{}^a \nabla^c \nabla^d R^{efgb}
= -\half R_{abcd} R_{efg}{}^a
( \nabla^c \nabla^d - \nabla^d \nabla^c )
R^{efgb}
\equn
$$
which is equivalent to $R^4$.
For $S_3$ we expand the first Riemann tensor using its cyclic symmetry
$$
\eqalign{
S_3 &= R_{abcd} \nabla^b R_{efg}{}^a \nabla^d R^{efgc}
\cr
&= - ( R_{acdb} + R_{adbc} ) \nabla^b R_{efg}{}^a \nabla^d R^{efgc}
\cr}
\equn
$$
By relabeling this is equal to
$$
\eqalign{
R_{acbd} \nabla^b R_{efg}{}^a ( \nabla^d R^{efgc} - \nabla^c R^{efgd} )
&= R_{acbd} \nabla^b R_{efg}{}^a ( \nabla^d R^{efgc} + \nabla^c R^{efdg} )
\cr
&= - R_{acbd} \nabla^b R_{efg}{}^a \nabla^g R^{efcd}
\cr}
\equn
$$
Taking the middle term out and noting that it multiplies a term which is
antisymmetric under exchange of $ab$
$$
\eqalign{
S_3 &= - \nabla^b R_{efg}{}^a R_{acbd} \nabla^g R^{efcd}
\cr
&= -\half ( \nabla^b R_{efg}{}^a - \nabla^a R_{efg}{}^b ) R_{acbd} \nabla^g R^{efcd}
\cr
&= +\half \nabla_g R_{ef}{}^{ab} R_{acbd} \nabla^g R^{efcd}
\cr
&= +\half  R_{acbd} \nabla_g R_{ef}{}^{ab} \nabla^g R^{efcd}
\cr}
\equn
$$
Integrating by parts
$$
\hskip 2.25cm
= -\half \nabla_g R_{acbd} R_{ef}{}^{ab} \nabla^g R^{efcd}
  -\half R_{acbd} R_{ef}{}^{ab} \nabla^2 R^{efcd}
\equn
$$
Taking the first term and integrating by parts with respect to the second $g$
$$
-\half \nabla_g R_{acbd} R_{ef}{}^{ab} \nabla^g R^{efcd} =
\half \nabla_g R_{acbd} \nabla^g R_{ef}{}^{ab} R^{efcd}
+\half \nabla^2 R_{acbd} R_{ef}{}^{ab} R^{efcd}
\equn
$$
The leading term is merely a relabeling of the original so that
$$
-\half \nabla_g R_{acbd} R_{ef}{}^{ab} \nabla^g R^{efcd} =
\frac{1}{4} \nabla^2 R_{acbd} R_{ef}{}^{ab} R^{efcd}
\equn
$$
and so
$$
S_3 = \frac{1}{4} \nabla^2 R_{acbd} R_{ef}{}^{ab} R^{efcd}
- \half R_{acbd} R_{ef}{}^{ab} \nabla^2 R^{efcd}
\equn
$$
Since the $\nabla^2 R_{abcd}$ leads to a combination of
$R^2$ tensors and derivatives acting upon Ricci tensors
then
the term $S_3$, for external graviton states, is equivalent to $R^4$ tensors.

For the last tensor,
$$
\eqalign{
S_4 =
R_{abcd} R^a{}_{efg} \nabla^g \nabla^d R^{becf}
&\equiv
-\nabla^g R_{abcd} R^a{}_{efg} \nabla^d R^{becf}
\cr
&= -\half \nabla^g R_{abcd} R^a{}_{efg} ( \nabla^d R^{becf} - \nabla^c R^{bedf} )
\cr
&= -\half \nabla^g R_{abcd} R^a{}_{efg} ( \nabla^d R^{becf} + \nabla^c R^{befd} )
\cr
&= +\half \nabla^g R_{abcd} R^a{}_{efg} \nabla^f R^{bedc}
\cr
&= -\half S_2
\cr}
\equn
$$
hence we can drop $S_4$ also.
Thus we are led to the conclusion that, for pure graviton amplitudes,
infinities can be removed by the introduction of purely $R^4$ counterterms.

From~\cite{Fulling}
the general $R^4$ counterterm is
$$
\frac{i}{(4 \pi)^4 \eps} \, \bigg[
a_1 T_1 + a_2 T_2 +a_3 T_3 +a_4 T_4 + a_5 T_5 +a_6 T_6 +a_7 T_7
\bigg]
\equn$$
where
$$
\eqalign{
T_1 &= ( R_{abcd} \, R^{abcd})^2
\cr
T_2 &= R_{abcd} \, R^{abc}{}_e \, R_{fgh}{}^d \, R^{fghe}
\cr
T_3 &= R^{ab}{}_{cd} \, R^{cd}{}_{ef} \, R^{ef}{}_{gh} \, R^{gh}{}_{ab}
\cr
T_4 &= R_{abcd} \, R^{ab}{}_{ef} \, R^{ce}{}_{gh} \, R^{dfgh}
\cr
T_5 &= R_{abcd} \, R^{ab}{}_{ef} \, R^c{}_g{}^e{}_h \, R^{dgfh}
\cr
T_6 &= R_{abcd} \, R^a{}_e{}^c{}_f \, R^e{}_g{}^f{}_h \, R^{bgdh}
\cr
T_7 &= R_{abcd} \, R^a{}_e{}^c{}_f \, R^e{}_g{}^b{}_h \, R^{fgdh}
\cr}
\equn$$
Additionally
the combination
$$
-{ T_1 \over 16}
+{T_2}
-{T_3  \over  8}
-T_4+2T_5-T_6 +2T_7
\equn$$
vanishes on-shell due to it
being proportional to the Euler form
$$
E \sim
\eps_{a_1a_2a_3a_4a_5a_6a_7a_8}
\eps^{b_1b_2b_3b_4b_5b_6b_7b_8}
R^{a_1a_2}{}_{b_1b_2}
R^{a_3a_4}{}_{b_3b_4}
R^{a_5a_6}{}_{b_5b_6}
R^{a_7a_8}{}_{b_7b_8}
\equn$$

The $R^4$ tensors are an interesting set. In $D=4$ they
degenerate into two independent tensors. One of these, the
famous ``Bel-Robinson'' tensor~\cite{BelRob}
was shown to be consistent with
supersymmetry and thus became a candidate counterterm for supergravity
theories \cite{R4sugra}. 
In higher dimensions the Bel-Robinson tensor
extends to a two-parameter set~\cite{DS}. For maximal supergravity theories the
uniqueness of the $R^4$ tensor extends to higher dimensions and is often
written
$$
t_8 \, t_8 \, R^4
\equn$$
where $t_8$ may be found in ref~\cite{GSW} eq.~(9.A.18). 
In $D=8$, in $N=2$ supergravity theory the four-point amplitudes
are exactly proportional to this tensor~\cite{GSB,DimShift}
and this tensor appears in the low-energy effective action of string theory~\cite{R4string}.
For $N=1$ supergravity there is an further combination
consistent with supersymmetry which appears if we calculate the
$N=1$ supergravity counterterms \cite{DJST}. It is interesting to calculate
the counterterms for simple gravity as a probe for the symmetries of the
gravitational theory.

Calculation with a general counterterm gives, for example,
$$
\eqalign{
\Mcount_1(1_g^{++},2_g^{++},3_g^{++},4_g^{++}) &=
8 \, (8 a_1 + 2 a_2 + 4 a_3 + a_6)
 { (s^2 + t^2 + u^2 )^2 \over s^2 u^2  }
 \times K_1 
\cr
\Mcount_{17}(1_g^{+I},2_g^{-I},3_g^{JK},4_g^{JK}) &=
 -\frac{s}{8} \, \left[ (4 a_2 - a_5 + 2 a_6) \, (t^2  + u^2 )
+ 2 \, (4 a_2 - a_5 - a_7)  t u  \right]\times K_{17}
\cr}
\equn$$
Clearly in this case it is not sufficient to look at amplitudes where
the external polarisations are four dimensional.
However, just from the $\Mloop(1_g^{++},2_g^{++},3_g^{++},4_g^{++})$ we can
clearly see that the counterterm does not vanish - although we can only
impose a single relationship between the coefficients of the six
counterterms.

In Appendix~C we calculate the infinities present in a sufficiently large class to determine the coefficients of the $T_i$
and
the $D=8$ counterterm for a real scalar loop is
$$
a_1= {11\over 29030400 } \;\;
a_2= {1\over 362880 } \;\;
a_3= {1\over 14515200 } \;\;
a_4=0 \;\;
a_5=0 \;\;
a_6= -{1\over 1814400 } \;\;
a_7= {1\over 453600 }
\equn
$$
The coefficient $a_4$ has been set to zero by choice but $a_5=0$ is
non-trivial.

We have also calculated the $R^4$ counterterms generated by other
types of matter circulating within the loop. Unlike the $D=6$ case,
the counterterms from different matter combinations are not simply
related. We present the coefficients of the $R^4$ counterterms
necessary to eliminate all divergence in four external graviton
amplitudes in table~\ref{GTA1-26}.  We have included the counterterms
where supersymmetric multiplets circulate for comparison.  (These have
been presented previously in ref~\cite{DJST}.) In $D=8$ the spinor
has eight degrees of freedom and both $N=1$ and $N=2$ supergravity
exist where the $N=2$ is the reduction of $D=10,N=2$
supergravity~\cite{CJS,ExtendedSugra}.  For $N=1$ there is the graviton
multiple and the matter multiplet. We have chosen to present the
combination of the two multiplets corresponding to the reduction of
$D=10,N=1$ supergravity (denoted $N=1^*$).  This prior to reduction
has particle content, in representations of $SO(8)$, $\ul{8}_c \otimes
(\ul{8}_v \oplus \ul{8}_s)$.  We have chosen to give the combined
contribution of a graviton and antisymmetric tensor in the loop
because this combination arise most naturally in superstring inspired
theories. (And in fact it is difficult to separate the two contributions in
string theory.)
In general one can rearrange the counterterms by addition of the Gauss
term and we have used this freedom to set $a_4=0$. For the $N=2$ case
the counterterm can be simplified
to 
$$
-{1 \over 64} \left( T_4 -4T_7 \right)
\equn
$$

\renewcommand{\thetable}{5.\arabic{subsection}}
\setcounter{subsection}{1}  

%
\begin{table}[!ht]
\hbox{
\def\tend{\cr
height9pt    &  & & & & & & & &   \cr
\noalign{ \hrule}
}
\def\t#1{\tilde{#1}}
\def\tw{\theta_W}
\vbox{\offinterlineskip
{
\hrule
\halign{
        &  \vrule #
        &\strut\hfil #\hfil\vrule
        &\strut\hfil #\hfil\vrule
        &\strut\hfil #\hfil\vrule
        &\strut\hfil #\hfil\vrule
        &\strut\hfil #\hfil\vrule
        &\strut\hfil #\hfil\vrule
        &\strut\hfil #\hfil\vrule
        &\strut\hfil #\hfil\vrule
       \cr
height12pt  & {\bf Matter}  
& $\;\;\;\;\;a_1\;\;\;\;\;$  
& $\;\;\;\;\;a_2\;\;\;\;\;$ 
& $\;\;\;\;\;a_3\;\;\;\;\;$  
& $\;\;\;\;\;\;\;\;a_4\;\;\;\;\; $  
& $\;\;\;\;\;a_5\;\;\;\;\; $   
& $\;\;\;\;\;a_6\;\;\;\;\;$ 
& $\;\;\;\;\; a_7\;\;\;\;\; $ & \tend
height18pt  & 
$\phi$  & 
${11\over 29030400}  $  &
${1\over 362880 } $  & 
${1\over 14515200 } $  & 
$0$  & 
$0 $  & 
$-{1\over 1814400 } $  & 
${1\over 453600 }  $  & 
\tend
%
height15pt  &
${\lambda }$  &
$-{89 \over 14515200   }$  &
${41 \over 725760  }$  &
$-{ 23 \over 907200 }$  &
${0   }$  &
${1 \over 20160 }$  &
$-{13 \over 453600 }$  &
${223 \over 1814400}  $  &
\tend
height15pt  &
$A_{a}$  &
$-{949 \over 77414400}  $  &
${29 \over 322560 }$  &
$-{389 \over 38707200} $  &
$0$  &
$-{1 \over 161280}$  &
${13 \over 1612800 } $  &
${31 \over 345600}  $  &
\tend
height15pt &
$B_{ab}+g_{ab} $  &
${3799\over 11612160} $  &
$-{841\over 145152}$  &
${2939\over 5806080} $  &
$0$  &
$-{521\over 40320}$  &
${5251\over 725760} $  &
${5779\over 362880}  $
\tend
height15pt  &
$\psi_{a}$  &
${103\over 414720}  $  &
$-{2867\over 725760} $  &
$+{457\over 725760} $  &
$0$  &
$-{31\over 4032}$  &
${139\over 45360 } $  &
$+{2143\over 362880}  $  &
\tend
%
height15pt  &
$N=2$  &
${1 \over 1024}  $  &
$-{1 \over 64} $  &
$ {1\over 512}$  &
${0}  $  &
$-{1 \over 32}$  &
${1 \over 64}$  &
${1 \over 32 } $  &
\tend
height15pt  &
$N=1^*$  &
$-{13 \over 737280}$  &
${7 \over 46080}$  &
$-{13 \over 368640 }$  &
${0 } $  &
${1 \over 23040 }$  &
$-{1 \over 46080 } $  &
${1 \over 4608 }  $  &
\tend
}
}
}
}
\nobreak
\caption[]{ Counterterms in $D=8$ for Various Matter Contents within the Loop
\label{GTA1-26}
\small
\smallskip}
\end{table}

The form of the counterterms, 
$$
\nabla^n R^m \hskip 2cm  {\rm where} \   n+2m = (D-2)L+2
\equn
$$
is symmetric under $(D-2) \leftrightarrow L $.  Which means for
example the form of the counterterms at $D=8,L=1$ is the same as that
for $D=4,L=3$ (up to dimensional dependent degeneracies).  It could be
hoped that studying counterterms in $D=8,L=1$ will provide information
about the $D=4,L=3$ case.  In fact for the $D=6,L=1$ case there does
appear to be a correlation of information - the $D=6,L=1$ counterterms
vanish for a supersymmetric theory as is the case for $D=4,L=2$.  For
the $D=4,L=3$ case the situation is far from clear. The unitarity
based results of ref.~\cite{BDDPR} indicate that for maximal
supergravity the three-loop amplitude in $D=4$ is finite.  This
conclusion has been supported by some field theoretical
evidence~\cite{HoweStelle}. We can examine the $D=8,L=1$ counterterms
to see if any understanding of the $D=4,L=3$ result can be obtained.  The
$D=8$ counterterms are written in terms of the six independent $T_i$
with non-zero coefficients. The combination compatible with maximal
supersymmetry appears to be unique~\cite{BDDPR,DS}.  If we were to
write this complete tensor in $D=4$ it is conceivable that it could
vanish in which case the vanishing of the infinity of the counterterm
would be a residual effect of reduction - analogous to the arguments
of ref~\cite{HoweStelle}. However the $D=8$ combination reduces to a
non-vanishing tensor - as evidence by the fact that the amplitude
$\Mcount(1^{--}_g,2^{--}_g,3^{++}_g,4^{++}_g)$ receives non-zero contributions
from this tensor so that the vanishing in the $D=4,L=3$
infinity remains a puzzle from this viewpoint.

\section{Conclusions}

In this paper we have used an extension of four dimensional helicity
to organise the scattering amplitudes for theories involving gravity.
This allows scattering amplitudes to be split into minimal physical
pieces which are generally simplier than the full amplitude. For many
purposes, such as determining the coefficients of counterterms, we
need only the results for a few helicity amplitudes. The individual
helicity amplitudes are physical and can be useful for testing
hypothesis and so we have included in our appendices rather more
calculations than we needed so that they may serve as a database for
others.

We have used the infinities in the physical four-point amplitudes to
determine the counterterm structure in $D=6,8$.  In the $D=6$ case the
counterterm was proportional to $N_B-N_F$ and hence vanishes in a theory with
equal numbers of bosonic and fermionic degrees of freedom such as a
supersymmetric theory. In contrast, the situation in $D=8$ is quite
different. The counterterms induced by different particles are
different and although the $N=2$
supersymmetric combinations are relatively simple they do not vanish.

Our investigations give no indications that a finite field theory of
gravity is possible. However our calculations should provide
indicators of the form of the low-energy effective action of the
fundamental theory of which gravity is merely the low-energy limit.

\newpage

\appendix

\renewcommand{\theequation}{A.\arabic{equation}}
\renewcommand{\thetable}{A.\arabic{table}}
\renewcommand{\thesection}{A.\arabic{section}}
\renewcommand{\thesubsection}{A.\arabic{section}}
\setcounter{equation}{0}  
\setcounter{section}{1}  
\setcounter{table}{0}  

\section*{Appendix A: Graviton Tree Amplitudes}  

In this appendix we present all the partial tree amplitudes for four
graviton scattering for dimension $D \geq 4$ in table~\ref{GTA}.
The partial amplitudes  are  given by
$$
\CMtree_i
= i \, \kappa^2 \, K_i \times F_i
$$
where the $K_i$ are products of spinor helicity factors.  When evaluating
the tree amplitude only the modulus of $K_i$ is relevant, however, the
complex phases are needed when trees are interfered with loops.  The
full form of the $K_i$ is given in table~\ref{Kfactors} and these will also
be used for the loop calculations.  For dimensions less than eight not all
the tree amplitudes exist, for example in $D=4$ only $M_1$, $M_2$ and $M_3$
exist. The minimal dimension that an amplitude exists in we call $D_M$
and this is also given in table~\ref{GTA}.

In general amplitudes are polynomials in $\eps\cdot k$ and
$\eps\cdot\eps'$.  In choosing the spinor helicity factors we
generally evaluate these from the highest order terms in $\eps\cdot
k$.  For four-point amplitudes, if we choose the spinor helicity
reference momentum, $q_i$, of an external states to be the external momentum
of another external state, this highest term has a unique form.  For
example, for $M_2({1}_g^{--},{2}_g^{++},{3}_g^{++},{4}_g^{++})$
we could choose the spinor helicity for the $\eps^{--}_{ab}$ to be $k_4$,
that is $\eps^{--}_{ab}(k_1;k_4)$. This means that
$$
\eqalign{
\eps^-(k_1;k_4) \cdot k_4 &= 0
\cr
\eps^-(k_1;k_4) \cdot k_3 &= -\eps^- (k_1;k_4)\cdot k_2
\cr}
$$
and the
leading term can be reduced to having a factor of $\eps_1^- \cdot k_2$.
This means the leading polynomial in $M_2({1}_g^{--},{2}_g^{++},{3}_g^{++},{4}_g^{++})$,
if we choose the four reference momenta $(q_1,q_2,q_3,q_4)=(k_4,k_1,k_1,k_1)$
will have a factor of
$$
\bigg(\eps_1^- \cdot k_2 \,
\eps_2^- \cdot k_3 \,
\eps_3^- \cdot k_2 \,
\eps_4^- \cdot k_2 \bigg)^2
$$
This is the K-factor for $M_2$ which can be reduced to spinor products
as given in table~\ref{Kfactors}.  For amplitudes such as
$M_{20}({1}_g^{--},{2}_g^{++},{3}_g^{IJ},{4}_g^{IJ})$, where $\eps_4 \cdot k_i=0$ for all
external momenta $k_i$, the highest order term ( after chooses, for
example, $(q_1,q_2)=(k_2,k_1)$ ) will be
$$
\left( \eps_1^- \cdot k_3 \,
\eps_2^+ \cdot k_3 \right)^2
$$
which is the $K$-factor.  The K factors are dependent on the choice of
reference momenta although the combination $K_i \times F_i$ is not.

\font\tenrm=cmr10
\font\eleven=cmr11
\font\twelverm=cmr12

\ninerm

%
\begin{table}[!ht]
\hbox{
\def\tend{\cr \noalign{ \hrule}}
\def\t#1{\tilde{#1}}
\def\tw{\theta_W}
\vbox{\offinterlineskip
{
\hrule
\halign{
       &  \vrule#
        &\strut\hfil #\hfil\vrule
        &\strut\hfil #\hfil\vrule
        &\strut\hfil #\hfil\vrule
        &\strut\hfil #\hfil\vrule
        &\strut\hfil #\hfil\vrule
        &\strut\hfil #\hfil\vrule
        &\strut\hfil #\hfil\vrule
        &\strut\hfil #\hfil\vrule
       \cr
height12pt  &{\bf Amplitude}  & $D_M$   & $F_i$ & $16 |K_i|$
 &{\bf Amplitude}  & $D_M$   & $F_i$ & $16 |K_i|$ & \tend
height12pt  & M1 $({}^{++},{}^{++},{}^{++},{}^{++})$ &
$4 $  &
$0 $  &
$s^2 u^2 $  &
M36$({}^{++},{}^{-I},{}^{+I},{}^{II})$  &
$5 $  &
$0 $  &
$4s u $  &
\tend
height12pt  & M2$({}^{--},{}^{++},{}^{++},{}^{++})$  &
$4 $  &
$0 $  &
$s^2 u^2 $  &
 M37 $({}^{++},{}^{+I},{}^{+I},{}^{JJ})$  &
$6 $  &
$0 $  &
$4s u $  &
\tend
height12pt  &M3 $({}^{--},{}^{--},{}^{++},{}^{++})$  &
$4 $  &
$4 { 1  \over stu } $  &
$s^4 $  &
 M38$({}^{--},{}^{+I},{}^{+I},{}^{JJ})$  &
$6 $  &
$-{ \sqrt{2} t \over \sqrt{3}su } $  &
$4s u $  &
\tend
height12pt  & M4$({}^{++},{}^{++},{}^{+I},{}^{+I})$ &
$5 $  &
$0$  &
$2 s u^2 $  &
M39 $({}^{++},{}^{-I},{}^{+I},{}^{JJ})$ &
$6 $  &
$0 $  &
$s^2u^2 $  &
\tend
height12pt  & M5$({}^{--},{}^{++},{}^{+I},{}^{+I})$  &
$5 $  &
$0 $  &
$2 s u^2 $  &
 M40 $({}^{+I},{}^{+J},{}^{IJ},{}^{II})$ &
$6 $  &
$0 $  &
$8s $  &
\tend
height12pt  & M6 $({}^{--},{}^{--},{}^{+I},{}^{+I})$  &
$5 $  &
$0 $  &
$2 s^3  $  &
 M41 $({}^{-I},{}^{+J},{}^{IJ},{}^{II})$ &
$6 $  &
${ 1\over 2 \sqrt{3} } $  &
$8s $  &
\tend
height12pt  & M7$({}^{++},{}^{++},{}^{+I},{}^{-I})$  &
$5 $  &
$0 $  &
$2s^2 u $  &
 M42$ ({}^{+I},{}^{+J},{}^{IJ},{}^{KK})$ &
$7 $  &
$0 $  &
$8s $  &
\tend
height12pt  & M8$ ({}^{--},{}^{++},{}^{-I},{}^{+I})$  &
$5 $  &
$2  \, { 1  \over su } $  &
$2  u^3 $  &
 M43$({}^{-I},{}^{+J},{}^{IJ},{}^{KK})$  &
$7 $  &
${ 1\over 2 \sqrt{3} } $  &
$8s $  &
\tend
height12pt  & M9 $({}^{+I},{}^{+I},{}^{+I},{}^{+I})$ &
$5 $  &
$0 $  &
$4 s u $  &
M44 $({}^{++},{}^{IJ},{}^{IJ},{}^{II})$ &
$6 $  &
$0 $  &
$8{su \over t} $  &
\tend
height12pt  & M10 $({}^{-I},{}^{+I},{}^{+I},{}^{+I})$ &
$5 $  &
$0 $  &
$4 t u  $  &
 M45 $({}^{++},{}^{IJ},{}^{IJ},{}^{KK})$ &
$7 $  &
$0 $  &
$8{su \over t} $  &
\tend
height12pt  & M11 $({}^{-I},{}^{-I},{}^{+I},{}^{+I})$ &
$5 $  &
$  { (2 t^2  + t u + 2 u^2 )\over 2stu } $  &
$4 s^2 $  &
 M46 $({}^{IJ},{}^{JK},{}^{KI},{}^{II})$ &
$7 $  &
${ u \over 2\sqrt{3} } $  &
$16 $  &
\tend
height12pt  & M12$ ({}^{+I},{}^{+I},{}^{+J},{}^{+J})$ &
$6 $  &
$0 $  &
$4 s u $  &
 M47 $({}^{IJ},{}^{JK},{}^{KI},{}^{LL})$  &
$8 $  &
$0 $  &
$16 $  &
\tend
height12pt  & M13$({}^{-I},{}^{+I},{}^{+J},{}^{+J})$  &
$6$  &
$0 $  &
$4tu  $  &
M48 $({}^{++},{}^{++},{}^{II},{}^{II})$ &
$5 $  &
$0 $  &
$4 s^2  $  &
\tend
height12pt  & M14 $({}^{-I},{}^{-I},{}^{+J},{}^{+J})$ &
$6 $  &
$-  {1 \over2  s} $  &
$4s^2  $  &
M49  $({}^{--},{}^{++},{}^{II},{}^{II})$ &
$5 $  &
${ 8 tu \over 3 s^3}  $  &
$4 s^2 $  &
\tend
height12pt  & M15 $({}^{-I},{}^{+I},{}^{-J},{}^{+J})$ &
$6 $  &
$ {(t - u) \over 2 su}$  &
$4 u^2 $  &  
M50 $({}^{++},{}^{++},{}^{II},{}^{JJ})$  &
$6 $  &
$0$  &
$4 s^2u^2 $  &
\tend
height12pt  & M16 $({}^{++},{}^{+I},{}^{+J},{}^{IJ})$ &
$6 $  &
$0 $  &
$4su $  &  M51  $({}^{--},{}^{++},{}^{II},{}^{JJ})$  &
$6 $  &
${4 t u \over 3 s^3}  $  &
$4 s^2  $  &
\tend
height12pt  & M17 $({}^{--},{}^{+I},{}^{+J},{}^{IJ})$  &
$6 $  &
$- { t \over\sqrt{2} su }$  &
$4s u $  &  M52 $({}^{+I},{}^{+I},{}^{II},{}^{II})$  &
$5 $  &
$0 $  &
$8s $  &
\tend
height12pt  & M18 $({}^{++},{}^{-I},{}^{+J},{}^{IJ})$ &
$6 $  &
$0 $  &
${ 2tu^2 \over s} $  &
  M53 $({}^{-I},{}^{+I},{}^{II},{}^{II})$  &
$5 $  &
${4(2 t^2  + 3 t u + 2 u^2 )\over 3s^2} $  &
$8s  $  &
\tend
height12pt  & M19  $({}^{++},{}^{++},{}^{IJ},{}^{IJ})$ &
$6 $  &
$0 $  &
$4s^2 $  &
 M54 $({}^{+I},{}^{+I},{}^{II},{}^{JJ})$  &
$6 $  &
$0 $  &
$8s $  &
\tend
height12pt  & M20 $({}^{--},{}^{++},{}^{IJ},{}^{IJ})$  &
$6 $  &
$ {s \over tu}  $  &
$4{t^2u^2 \over s^2}  $  &
 M55 $({}^{-I},{}^{+I},{}^{II},{}^{JJ})$  &
$6 $  &
${ 3 t^2  + 4 t u + 3 u^2 \over 3s^2}  $  &
$8s $  &
\tend
height12pt  & M21  $({}^{+I},{}^{+I},{}^{IJ},{}^{IJ})$  &
$6 $  &
$0 $  &
$8s $  &  M56 $({}^{+I},{}^{+I},{}^{JJ},{}^{JJ})$  &
$6 $  &
$0 $  &
$8s$  &
\tend
height12pt  & M22 $({}^{+I},{}^{-I},{}^{IJ},{}^{IJ})$  &
$6 $  &
$ { (t^2+u^2) \over 4 tu} $  &
${8t u \over s}  $  &
 M57 $({}^{-I},{}^{+I},{}^{JJ},{}^{JJ})$  &
$6 $  &
${ (t-u)^2 \over 3s^2}  $  &
$8s $  &
\tend
height12pt  & M23 $({}^{+I},{}^{+I},{}^{JK},{}^{JK})$  &
$7 $  &
$0 $  &
$8s $  &  M58 $({}^{+I},{}^{+I},{}^{JJ},{}^{KK})$  &
$7 $  &
$0 $  &
$8s $  &
\tend
height12pt  & M24 $({}^{+I},{}^{-I},{}^{JK},{}^{JK})$  &
$7 $  &
$-\frac{1}{2} \,  $  &
${8tu\over s} $  &
 M59 $({}^{-I},{}^{+I},{}^{JJ},{}^{KK})$  &
$7 $  &
${ t^2 +u^2 \over 3s^2}  $  &
$8s $  &
\tend
height12pt  & M25 $({}^{+I},{}^{+J},{}^{IK},{}^{JK})$  &
$7 $  &
$0 $  &
$8s  $  & M60 $({}^{IJ},{}^{IJ},{}^{II},{}^{II})$  &
$6 $  &
$-{ 2 t^2  + t u + 2 u^2 ) \over 3s } $  &
$16 $  &
\tend
height12pt  & M26 $({}^{+I},{}^{-J},{}^{IK},{}^{JK})$  &
$7 $  &
$ -{s \over 4 u} $  &
${8tu \over s} $  &  
M61 $({}^{IJ},{}^{IJ},{}^{II},{}^{JJ})$  &
$6 $  &
${ (2 t + u) (t + 2 u) \over 3s }  $  &
$16 $  &
\tend
height12pt  & M27 $({}^{++},{}^{IJ},{}^{JK},{}^{KI})$  &
$7 $  &
$0 $  &
${8 t u \over s}  $  &
 M62 $({}^{++},{}^{II},{}^{II},{}^{II})$ &
$5 $  &
$0  $  &
$8 {s u\over t}  $  &
\tend
height12pt  & M28 $({}^{IJ},{}^{IJ},{}^{IJ},{}^{IJ})$  &
$6 $  &
$\,  \, { (s^2 + t^2 + u^2 )^2 \over 16 stu} $  &
$16$  & 
M63 $({}^{++},{}^{II},{}^{II},{}^{JJ})$ &
$6$  &
$0 $  &
$8 {s u\over t} $  &
\tend
height12pt  & M29 $({}^{IJ},{}^{IJ},{}^{IK},{}^{IK})$  &
$7 $  &
$- {(t^2+u^2) \over 8 s}  $  &
$16$  &  M64 $({}^{++},{}^{II},{}^{JJ},{}^{KK})$ &
$7 $  &
$0 $  &
$8 {s u\over t} $  &
\tend
height12pt  & M30 $({}^{IJ},{}^{IJ},{}^{KL},{}^{KL})$  &
$8$  &
$ { t u \over 4 s} $  &
$16 $  & M65 $({}^{II},{}^{II},{}^{II},{}^{II})$ &
$5 $  &
$\frac{(s^2+t^2+u^2)^2}{stu}$  &
$16 $  &
\tend
height12pt  & M31 $({}^{IJ},{}^{JK},{}^{KL},{}^{LI})$  &
$8 $  &
$\frac{u}{8}  $  &
$16 $  &  M66 $({}^{II},{}^{II},{}^{II},{}^{JJ})$ &
$6 $  &
$\frac{1}{3} \, \frac{(s^2+t^2+u^2)^2}{stu} $  &
$16 $  &
\tend
height12pt  & M32 $({}^{++},{}^{++},{}^{++},{}^{II})$ &
$5 $  &
$0 $  &
$2stu $  & M67 $({}^{II},{}^{II},{}^{JJ},{}^{JJ})$ &
$6 $  &
$\frac{4(t^4  + 2 t^3  u + 7 t^2  u^2  + 2 t u^3  + u^4 )}{9stu} $  &
$16 $  &
\tend
height12pt  & M33 $({}^{--},{}^{++},{}^{++},{}^{II})$  &
$5 $  &
$0 $  &
$2stu $  & M68 $({}^{II},{}^{II},{}^{JJ},{}^{KK})$ &
$7 $  &
$\frac{4(t^4  + 2 t^3  u + 5 t^2  u^2  + 2 t u^3  + u^4 )}{9stu} $  &
$16 $  &
\tend
height12pt  &M34 $({}^{++},{}^{+I},{}^{+I},{}^{II})$  &
$5 $  &
$0 $  &
$4s u $  &  M69 $({}^{II},{}^{JJ},{}^{KK},{}^{LL})$ &
$8 $  &
$\frac{1}{9} \, \frac{(s^2+t^2+u^2)^2}{stu} $  &
$16 $  &
\tend
height 12pt  & M35 $({}^{--},{}^{+I},{}^{+I},{}^{II})$ &
$5 $  &
$-{ \sqrt{6} t \over s u}  $  &
$4s u $  &  &
  &
  &
  &
\tend
}
}
}
}
\nobreak
\caption[]{Graviton Tree Amplitudes
\label{GTA}
\small
\smallskip}
\end{table}

%
\begin{table}[!ht]
\hbox{
\def\tend{\cr \noalign{ \hrule}}
\def\t#1{\tilde{#1}}
\def\tw{\theta_W}
\vbox{\offinterlineskip
{
\hrule
\halign{
       &  \vrule#
        &\strut\hfil #\hfil\vrule
        &\strut\hfil #\hfil\vrule
        &\strut\hfil #\hfil\vrule
        &\strut\hfil #\hfil\vrule
       \cr
height12pt  & \hskip 1cm $K_i$ \hskip 1cm  & \hskip 1.5 cm Value \hskip 1.5 cm  &
  \hskip 1cm $K_i$ \hskip 1cm   & \hskip 1.5 cm Value \hskip 1.5 cm &    \tend
height14pt  & $16 K_1$ &
$\left({ s^2 t u \over \spa1.2\spa2.3\spa3.4\spa4.1}\right)^2$  &
$4K_{36}$ &
$- { st\over u} { \spa4.2^2\spb1.3^2\over \spa4.1^2 } $  &
\tend
height12pt  & $16 K_2$ &
${ (\spa1.2\spb2.4\spb2.3)^4 / t^2 } $  &
$K_{37} $  &
$K_{16} $  &
\tend
height14pt  &$ 16K_3$  &
$\left( { st \spa1.2^4 \over \spa1.2\spa2.3\spa3.4\spa4.1 }\right)^2 $  &
$K_{38}$ &
$K_{35} $  &
\tend
height14pt  & $8K_4$ &
$-u^2 \left( { \spb1.2\spb2.3 \over \spa4.1 } \right)^2 $  &
$K_{39} $ &
$K_{36} $  &
\tend
height14pt  & $8K_5$ &
$-\left( {\spa1.2^2 \spb2.3^2 \spb2.4^4 \over\spb1.4^2 }\right)  $  &
$K_{40} $ &
$K_{21} $  &
\tend
height12pt  & $8K_6 $ &
$-\spa1.2^4 \spb3.4^2 $  &
$2 K_{41} $ &
$-s { \spa1.4\spb4.2 \over \spa2.4\spb4.1 }  $  &
\tend
height12pt  & $8K_7$ &
$-{ \spb1.2^4 \spb1.3^2 \spa4.1^2 / t} $  &
$K_{42}$  &
$K_{21} $  &
\tend
height12pt  & $8K_8$   &
$-{ \spa1.3^4 \spb2.3^2 \spb2.4^2 / t }  $  &
$K_{43}$  &
$K_{41} $  &
\tend
height14pt  & $K_9$  &
$\sqrt{K_1}  $  &
$2K_{44} $ &
$\left( { \spa4.2\spb2.1 \over\spa4.1 } \right)^2 $   &
\tend
height12pt  & $4 K_{10}$  &
 $ { (\spa1.2\spb2.3\spb2.4)^2 / s } $  &
$K_{45} $&
$K_{44} $  &
\tend
height12pt  & $K_{11}$  &
$\sqrt{K_3}  $  &
$K_{46}$  &
$1 $  &
\tend
height12pt  & $K_{12}$ &
 $\sqrt{K_1}  $ &
$K_{47} $ &
$1 $  &
\tend
height12pt  & $K_{13}$ &
$K_{10}$  &
$K_{48} $ &
$K_{21}^2 $  &
\tend
height12pt  & $K_{14} $ &
$\sqrt{K_3} $  &
$K_{49}$&
$K_{41}^2 $  &
\tend
height14pt  & $4 K_{15} $ &
$-\left( { \spa1.3^4 st \over \spa1.2\spa2.3\spa3.4\spa4.1 }\right)$  &
$K_{50} $ &
$K_{21}^2 $  &
\tend
height12pt  & $4K_{16} $ &
$-\spb1.2^2\spb1.3^2  $  &
$K_{51} $ &
$K_{41}^2 $  &
\tend
height14pt  & $4K_{17} $ &
$-\left( { \spb4.2\spa2.1 \spb2.3\over \spb1.4 }\right)^2 $  &
$K_{52} $  &
$K_{21} $  &
\tend
height11pt  & $4K_{18} $ &
${ \spa2.3^2 \spb1.3^4 / s} $  &
$K_{53} $ &
$K_{41} $  &
\tend
height10pt  & $K_{19}  $ &
$K_{21}^2 $  &
$K_{54} $ &
$K_{21} $  &
\tend
height11pt  & $4K_{20} $  &
${ (\spb2.3\spa3.1 )^4/ s^2} $  &
$K_{55} $  &
$K_{41} $  &
\tend
height10pt  & $ 2 K_{21}  $ &
$-{ \spb1.2^2 } $  &
$K_{56} $  &
$K_{21} $  &
\tend
height11pt  & $2K_{22} $  &
$-{ ( \spa2.3\spb3.1 )^2/ s } $  &
$K_{57} $ &
$K_{41} $  &
\tend
height10pt  & $K_{23} $ &
$K_{21} $  &
$K_{58} $ &
$K_{21} $  &
\tend
height10pt  & $K_{24} $ &
$K_{22} $  &
$K_{59} $ &
$K_{41} $  &
\tend
height10pt  & $K_{25} $  &
$K_{21} $  &
$K_{60} $  &
$1 $  &
\tend
height10pt  & $K_{26} $ &
$K_{22} $  &
$K_{61} $  &
$1 $  &
\tend
height14pt  & $2K_{27} $ &
$\left( { \spa2.3 \spb3.1 \over \spa2.1 } \right)^2 $  &
$K_{62} $ &
$K_{44}$   &
\tend
height10pt  & $K_{28} $  &
$1 $  &
$K_{63}$  &
$K_{44} $  &
\tend
height10pt  &  $K_{29} $ &
$1$  &
$K_{64}$  &
$K_{44} $  &
\tend
height10pt  & $K_{30}$  &
$1$  &
$K_{65} $ &
$1 $  &
\tend
height10pt  & $K_{31}$  &
$1 $  &
$K_{66} $ &
$1 $  &
\tend
height12pt  & $8K_{32} $ &
$(  \spb1.2\spb2.3\spb1.3 )^2  $  &
$K_{67} $ &
$1$ &
\tend
height12pt  & $8K_{33} $  &
$ (  \spa1.4^2 \spb2.3 \spb3.4\spb2.4 )^2 / t^2 $  &
$K_{68}$  &
$1 $  &
\tend
height12pt  &$4K_{34} $ &
$-\spb1.2^2\spb1.3^2 $  &
$K_{69}$  &
$1 $  &
\tend
height 14pt  & $4K_{35} $ &
$-\left( {\spb4.2\spb2.3\spb2.1 \over \spb4.1} \right)^2 $  &
{} &
{} &
\tend
}
}
}
}
\nobreak
\caption[]{The K factors for the Graviton Amplitudes
\label{Kfactors}
\small
\smallskip}
\end{table}

\renewcommand{\theequation}{B.\arabic{equation}}
\renewcommand{\thesection}{B.\arabic{section}}
\renewcommand{\thesubsection}{B.\arabic{section}}
\setcounter{equation}{0}  
\setcounter{section}{1}  

\vfill\eject

\section*{Appendix B: Examples of the Cutkosky Cutting Technique}

\eleven

Here we  demonstrate the steps necessary to evaluate the
all-$\epsilon$ form of the one-loop 
amplitudes which involve both gravitons and two-forms. 
This also illustrates
the links between the graviton and form scattering.

To evaluate the two-particle cuts we need the tree amplitudes for two
external particles, with momenta in four dimensions, and for two
internal particles with momenta in $D$ dimensions, where
$D=2N-2\eps$.  Since we are examining amplitudes with complex
scalar loops
these two internal particles should be complex scalars.
The KLT relationships can be used to determine these ``primitive
amplitudes'' from the Yang-Mills amplitudes,
$$
\eqalign{M_4^{\rm tree}( 1,2,3,4)= -
\frac{i s_{12}}{4} \, A_4^{\rm tree}(1,2,3,4) \, A_4^{\rm tree}(1,2,4,3)
\cr}
$$
The Yang-Mills amplitude we shall need is
$$
\eqalign{
A_4^{\rm tree}(1_s,2^I,3^J,4_s)
= -2i \left(
{ L^IL^J \over \Box_{23} } -{ \Box_{32} \delta^{IJ} \over 2s_{23} }
\right)
\cr}
$$
where $\Box_{23}\equiv (L_1-k_2)^2$ is the propagator from leg two to
leg three in a clockwise manner.  From these we can deduce that
$$
\eqalign{
M^{\rm tree}(s;g^{IJ};g^{KL};s)
={ 1 \over \sqrt{2}^2 } & \Bigl(
 M^{\rm tree, P}(s;I,J;K,L;s)  +M^{\rm tree, P}(s;J,I;K,L;s)
\cr & \null \hskip 1.5 truecm
+M^{\rm tree, P}(s;I,J;L,K;s)  +M^{\rm tree, P}(s;J,I;L,K;s)
\Bigr)
\cr
=& {i s }\Bigl(  A^{\rm tree}(s,1^I,2^K,s)A^{\rm tree}(s,2^J,1^L,s)
+A^{\rm tree}(s,1^J,2^K,s)  A^{\rm tree}(s,2^I,1^L,s)
\Bigr)\cr
=&{is } \left(
 { L^IL^KL^JL^L \over \Box_{12}\Box_{21} }
+{ L^JL^KL^IL^L \over \Box_{12}\Box_{21} }
\right)
\cr
=&2is{ L^IL^JL^KL^L \over \Box_{12}\Box_{21} }
\cr
=&-2i{ L^IL^JL^KL^L }\left( {1 \over \Box_{12}}+{1 \over\Box_{21} } \right)
\cr}
$$
which will prove to be an extremely useful form when cutting.
We have used the identity
$$
{ 1 \over \Box_{12}\Box_{21}}
+{ 1 \over s_{22}\Box_{21}}
+{ 1 \over \Box_{12}s_{12}}
=0
$$
which follows from $s_{12}+\Box_{12}+\Box_{21}=0$, which is
true since the tree amplitude is fully on-shell.
Similarly,
$$
\eqalign{
M^{\rm tree}(s;g^{IJ};g^{IK};s)
&=- 2i L^IL^JL^IL^K
\left( {1 \over \Box_{12}}+{1 \over\Box_{21} } \right)
-{i  L^JL^K \over 2 }
\cr
M^{\rm tree}(s;g^{IJ};g^{IJ};s)
&=-{2i}L^IL^I L^JL^J
\left( {1 \over \Box_{12}}+{1 \over\Box_{21} } \right)
-{i \over 2} \Big(L^IL^I+L^JL^J \Bigr)
+{i \over 4s} \Box_{12}\Box_{21}
\cr}
$$

Amplitudes involving forms are obtained
from the same primitive amplitudes but with appropriate minus signs.
For Amplitudes with one $g$ and one $B$ the tree amplitude vanishes
$$
M^{\rm tree}(s;g^{IJ};B^{KL};s) =M^{\rm tree}(s;g^{IJ};B^{IK};s)
=M^{\rm tree}(s;g^{IJ};B^{IJ};s) = 0
$$
For amplitudes with two $2$-forms we obtain
$$
\eqalign{
M^{\rm tree}(s;B^{IJ};B^{KL};s)
&={ 1 \over \sqrt{2}^2 }
\Bigl(
M^{\rm tree, P}(s;I,J;K,L;s) -M^{\rm tree, P}(s;J,I;K,L;s)
\cr
& \null\hskip 1.5 truecm
-M^{\rm tree, P}(s;I,J;L,K;s) +M^{\rm tree, P}(s;J,I;L,K;s)
\Bigr)
\cr
&= 0
\cr
M^{\rm tree}(s;B^{IJ};B^{IK};s) &={ 1 \over \sqrt{2}^2 } \Bigl(
M^{\rm tree, P}(s;I,J;I,K;s) -M^{\rm tree, P}(s;J,I;I,K;s)
\cr
& \null\hskip 1.5 truecm
-M^{\rm tree, P}(s;I,J;K,I;s) +M^{\rm tree, P}(s;J,I;K,I;s) \Bigr)
\cr
&= {is }\Bigl( A^{\rm tree}(s,1^I,2^I,s)A^{\rm tree}(s,2^J,1^K,s)
-A^{\rm tree}(s,1^I,2^K,s)
A^{\rm tree}(s,2^J,1^I,s) \Bigr)
\cr
&= -{i  L^JL^K\over 2 }
\cr
M^{\rm tree}(s;B^{IJ};B^{IJ};s) &={ 1 \over \sqrt{2}^2 } \Bigl(
M^{\rm tree, P}(s;I,J;I,J;s) -M^{\rm tree, P}(s;J,I;I,J;s)
\cr
& \null\hskip 1.5 truecm
-M^{\rm tree, P}(s;I,J;J,I;s) +M^{\rm tree, P}(s;J,I;J,I;s) \Bigr)
\cr
&= {is }\Big( A^{\rm tree}(s,1^I,2^I,s)A^{\rm tree}(s,2^J,1^J,s)
-A^{\rm tree}(s,1^J,2^I,s)
A^{\rm tree}(s,2^I,1^J,s) \Bigr)
\cr &=-{i \over 2} \Big( L^IL^I+L^JL^J
\Bigr) +{i  \over 4s} \Box_{12}\Box_{21} \cr}
$$

We now
have the building blocks necessary to evaluate the cuts in
examples the one-loop amplitudes.

\vskip 0.3 truecm
\noindent
{\bf Example 1:} $\Mloop(g_1^{IJ};g_2^{IK}; B_{3}^{LJ}, B_4^{LK} )$

A four-point amplitude will have, in general, three cuts - in the $s$,
$t$ and $u$ invariants.  For this amplitude, to all orders in $\epsilon$ the $t$ and $u$
cuts vanish identically, since the tree amplitudes for these cuts vanish,
 and the amplitude only has a $s$-cut as given in
eq.~(4.2)
$$
\sum_{{\rm internal}\atop {\rm states,s}}
\left. \int {d^D L_1 \over (2 \pi)^D} \,
  {i\over L_1^2} \,
{ M}_{4}^\tree(-L_1^s,1,2,L_3^s)
\times
\,{i \over L_3^2}\,
{ M}_{4}^\tree (-L_3^s,3,4,L_1^s) \right|_{\rm s- cut}
$$
Manipulating the tree amplitudes within this cut
$$
\eqalign{
&\Mtree(s_{l_1}, g_1^{IJ};g_2^{IK}; s_{l_2})
\times
\Mtree(s_{l_2}; B_{3}^{LJ}, B_4^{LK} s_{l_1} )
\cr
& =
i^2 \left(- 2 L^IL^JL^IL^K
\left( {1 \over \Box_{12}}+{1 \over\Box_{21} } \right)
-{  L^JL^K \over 2 } \right)
\times \Bigl(-{  L^JL^K \over 2s }\Bigr)
\cr
& = -{  L^IL^JL^IL^K  L^JL^K\over s\Box_{12} }
- {  L^IL^JL^IL^K  L^JL^K\over s\Box_{21} }
- { L^JL^K  L^JL^K\over 4s }
\cr}
$$
and inserting this product into the cut, after adding the two propagators,
these three terms can be recognised as the
cuts of two triangle integrals and a bubble integral,
$$
-{i\over(4\pi)^{D/2} s}I_3^D[L^IL^JL^IL^K  L^JL^K]-
{i\over(4\pi)^{D/2} s}I_3^D[L^IL^JL^IL^K  L^JL^K]
+{i\over(4\pi)^{D/2} 4s} I^D_2[L^JL^K  L^JL^K]
$$
The effect of the internal momentum is to ``shift'' the dimension of the integral, for example,
$$
I^D[L^IL^I] ={ 1\over 2} I^{D+2}[1] \;,\;\;\;
I^D[L^IL^IL^JL^J] ={ 1\over 4} I^{D+4}[1] \;,\;\;\; etc
$$
 and this integral is equal to
$$ ={i \over (4\pi)^{D/2} } \left( -{1\over 4}   I_3^{D+6}(s)
+{1 \over 16} I_2^{D+4}(s)  \right)
$$
where we have chosen to indicate the momentum invariant upon which the
integrals depend. 
This expression has the correct value for all the cuts of the amplitude to all orders in $\epsilon$ so that
$$
\Mloop(g_1^{IJ};g_2^{IK}; B_{3}^{LJ}, B_4^{LK} )
={i \over (4\pi)^{D/2} }
\left( -{1\over 4}   I_3^{D+6}(s)
+{1 \over 16} I_2^{D+4}(s)  \right)
$$
The infinities in this amplitude match those of
table~\ref{InfinitiesGGBB} (after dividing by two to get the
contribution from a real scalar.) although the full one-loop
amplitude contains much more information than merely the ultra-violet infinities.

\vskip 0.3 truecm
\noindent
{\bf Example 2:}  $\Mloop(B_1^{IJ};B_2^{IK}; B_{3}^{LJ}, B_4^{LK} )$

For this amplitude the $t$-cut is identically zero leaving $s$ and $u$ cuts.
Firstly the $s$-cut gives
$$
\eqalign{
\Mtree(s_{l_1}, B_1^{IJ};B_2^{IK}; s_{l_2})
\times
\Mtree(s_{l_2}; B_{3}^{LJ}, B_4^{LK} s_{l_1} )
& = {i L^J L^K \over 2 } \times {i L^J L^K \over 2 }
\cr}
$$
After inserting this into the cut, we find that the $s$-cut will be
the cut of the bubble integral
$$
{ i \over 4(4\pi)^{D/2} }I_2^D[L^J L^K L^J L^K ]
$$
Again the internal momentum leads to a shifted integral,
$$
{i \over 16(4\pi)^{D/2} } I_2^{D+4}(s)
$$
The $u$-cut is identical, after substituting $s\rightarrow u$,
giving the total amplitude as
$$
\Mloop(B_1^{IJ};B_2^{IK}; B_{3}^{LJ}, B_4^{LK} )
= {i \over 16(4\pi)^{D/2} }
\left(
I_2^{D+4}(s)+I_2^{D+4}(u)
\right)
$$
whose infinities match those contained in table~\ref{BTA}.

\vskip 0.3 truecm
\noindent
{\bf Example 3:}  $\Mloop(g_1^{IJ};g_2^{IK}; g_{3}^{KL}, g_4^{JL} )$

This amplitude has cuts in all three channels, the simplest being the
$u$ channel where the cut is
$$
\eqalign{
\Mtree(s_{l_1}, g_1^{IJ}; g_3^{Kl}; s_{l_2})
& \times
\Mtree(s_{l_2}; g_{2}^{IK}; g_4^{LJ}; s_{l_1} )
\cr
&=
-2i L^IL^JL^KL^L \Bigl(
{1 \over \Box_{13}}+ {1 \over\Box_{31} } \Bigr)
\times
-2i { L^IL^JL^KL^L}\Bigl(
{1  \over \Box_{24}}+ {1 \over\Box_{42} }\Bigr)
\cr}
$$
which is the cut of
$$
{i\over 2(4\pi)^{D/2} } \Bigl(
I_{1243}^{D+8}(s,u)+I_{1423}^{D+8}(t,u)
\Bigr)
$$
The $s$-cut is then
$$
{is }\Big( 2{ L^IL^JL^IL^K \over \Box_{12}\Box_{21} }
-{  L^JL^K \over 2s }
\Bigr)
\times {
i s }\Big( 2{ L^LL^JL^LL^K \over \Box_{34}\Box_{43} }
-{  L^JL^K \over 2s }
\Bigr)
$$
which will be the cut of
$$
{i \over (4\pi)^{D/2} } \left(
{1\over 2} \Bigl(
I_{1243}^{D+8}(s,u)+I_{1234}^{D+8}(s,t)
\Bigr)
-{1\over 2}  I_3^{D+6}(s)
+{1 \over 16} I_2^{D+4}(s)
\right)
$$
The $t$-cut can be obtained from the $s$ by relabeling.
Putting the cuts together we find the entire amplitude is
$$\eqalign{
\Mloop(g_1^{IJ};g_2^{IK}; g_{3}^{KL}, g_4^{JL} ) =
{i \over (4\pi)^{D/2} } \biggl(
{1\over 2} \Bigl( &
I_{1234}^{D+8}(s,t)+I_{1243}^{D+8}(s,u)+
I_{1423}^{D+8}(t,u)
\Bigr)\cr
& -{1\over 2} \Bigl(  I_3^{D+6}(s)+ I_3^{D+6}(t)\Bigr)
+{1 \over 16} \left(I_2^{D+4}(s)+I_2^{D+4}(t) \right)
\biggr)
\cr}
$$

\vskip 1.0 truecm

\renewcommand{\theequation}{C.\arabic{equation}}
\renewcommand{\thetable}{C.\arabic{table}}
\renewcommand{\thesection}{C.\arabic{section}}
\renewcommand{\thesubsection}{C.\arabic{section}}
\setcounter{equation}{0}  
\setcounter{section}{1}  
\setcounter{table}{0}  

\section*{Appendix C:  Infinities in One-Loop Four Graviton Amplitudes}

\setcounter{equation}{0}  



We have calculated the infinities in the partial amplitudes for four
graviton scattering in $D=6,8,10$ for real scalar loops. In $D=6$ a single,
non-zero infinity will be enough to specify the coefficient of the
single counterterm required to make the four graviton amplitudes
finite. In $D=8$ the first ten amplitudes are sufficient to fix the
coefficients of the six $R^4$ tensors which can appear.
The infinities in table~\ref{GOLA} over-specify the system considerable and
all the infinities match the counterterms precisely.

The infinities in the loop amplitudes are
$$
\left. \CMloop_i (1_g,2_g,3_g,4_g) \right|_{1/\eps} =
 { i \kappa^4 \over (4\pi)^{D/2} \eps}
\times K_i
\times \Floop_i
$$
where the $\Floop_i$ are
given in the following table~\ref{GOLA}, labeled by their dimension.

\vfill\eject

\ninerm

\vskip 1cm

\begin{table}[!ht]
\hbox{
\def\tend{\cr \noalign{ \hrule}}
\def\t#1{\tilde{#1}}
\def\tw{\theta_W}
\vbox{\offinterlineskip
{
\hrule
\halign{
       &  \vrule#
        &\strut\hfil #\hfil\vrule
        &\strut\hfil #\hfil\vrule
        &\strut\hfil #\hfil\vrule
        &\strut\hfil #\hfil\vrule
        &\strut\hfil #\hfil\vrule
        &\strut\hfil #\hfil\vrule
       \cr
height12pt  &{\bf Amplitude}  & $D_M$  &
$D=6$ & $D=8$ & $D=10$ & $16 |K_i|$  &\tend
height12pt  & M1$({}^{++},{}^{++},{}^{++},{}^{++})$ &
$4 $  &
${t \over 504 su} $  &
${(s^2+t^2+u^2)^2 \over 15120 s^2u^2} $  &
${(s^2+t^2+u^2) \, t \over 31680 su} $  &
$s^2u^2 $  &
\tend
height12pt  & M2$({}^{--},{}^{++},{}^{++},{}^{++})$  &
$4 $  &
${t \over 5040 su}  $  &
$0 $  &
$0 $  &
$s^2u^2 $  &
\tend
height12pt  &M3$({}^{--},{}^{--},{}^{++},{}^{++})$  &
$4 $  &
$0 $  &
${1 \over 6300} $  &
${s \over 83160} $  &
$s^4 $  &
\tend
height12pt  & M4$({}^{++},{}^{++},{}^{+I},{}^{+I})$ &
$5 $  &
${t \over 2520 u} $  &
${s(2 t^2 + 3 tu + 2u^2) \over 30240 u^2} $  &
${10t^4+46t^3u+69t^2u^2+46tu^3+10u^4 \over 1995840 u^2} $  &
$2 su^2 $  &
\tend
height12pt  & M5$({}^{--},{}^{++},{}^{+I},{}^{+I})$  &
$5 $  &
$0 $  &
$0 $  &
$-{ t^2 \over 3326400} $  &
$2su^2 $  &
\tend
height12pt  & M6$({}^{--},{}^{--},{}^{+I},{}^{+I})$  &
$5 $  &
$0 $  &
${s \over 16800} $  &
${55t^2+116tu+55u^2 \over 9979200} $  &
$2 s^3 $  &
\tend
height12pt  & M7$({}^{++},{}^{++},{}^{+I},{}^{-I})$  &
$5 $  &
$-{t \over 10080 s} $  &
$0 $  &
$-{st \over 1995840} $  &
$2s^2u $  &
\tend
height12pt  & M8$ ({}^{--},{}^{++},{}^{-I},{}^{+I})$  &
$5 $  &
$0 $  &
${ t \over 50400} $  &
$ -{(12s+10t)t \over 19958400} $  &
$2u^3 $  &
\tend
height12pt  & M9$({}^{+I},{}^{+I},{}^{+I},{}^{+I})$ &
$5 $  &
$ {t \over 3360 }$  &
${ 19(s^2  + s t + t^2 )^2 \over 302400 s u} $  &
${11(s^2  + s t + t^2 )t \over 725760} $  &
$4su $  &
\tend
height12pt  & M10$({}^{-I},{}^{+I},{}^{+I},{}^{+I})$ &
$5 $  &
$0 $  &
$0 $  &
${ s (s^2+st+t^2)\over  39916800}$  &
$4{tu}  $  &
\tend
height12pt  & M11$({}^{-I},{}^{-I},{}^{+I},{}^{+I})$ &
$5 $  &
$0 $  &
${ 25 t^2  + 36 u t + 25 u^2 \over 604800 }$  &
${ s (35 u^2  + 86 u t + 35 t^2 )\over 13305600 }$  &
$4s^2 $  &
\tend
height12pt  & M12$ ({}^{+I},{}^{+I},{}^{+J},{}^{+J})$ &
$6 $  &
${t \over 10080} $  &
${25t^4 + 86t^3u +129t^2u^2 + 86tu^3 + 25u^4 \over 907200 s u} $  &
${20t^4 + 87t^3u +127t^2u^2 + 87tu^3 + 20u^4 \over 7983360 u}  $  &
$4su $  &
\tend
height12pt  & M13$({}^{-I},{}^{+I},{}^{+J},{}^{+J})$  &
$6$  &
$0 $  &
${s^2 \over 201600} $  &
${s \, (3t^2+11tu+3u^2) \over 39916800} $  &
$4tu $  &
\tend
height12pt  & M14$({}^{-I},{}^{-I},{}^{+J},{}^{+J})$ &
$6 $  &
$0 $  &
${47t^2+84tu+47u^2 \over 1814400} $  &
${s \, (35t^2 + 72 tu + 35u^2) \over 13305600} $  &
$4s^2 $  &
\tend
height12pt  & M15$({}^{-I},{}^{+I},{}^{-J},{}^{+J})$ &
$6 $  &
$0 $  &
${16t^2+22tu+17u^2 \over 1814400} $  &
$-{ 10 t^3  + 36 u t^2  + 36 u^2  t + 5 u^3 \over 39916800}$  &
$4u^2 $  &
\tend
height12pt  & M16$({}^{++},{}^{+I},{}^{+J},{}^{IJ})$ &
$6 $  &
$-{ t \sqrt{2} \over 20160}  $  &
${ \sqrt{2} (3s^2-4su+3u^2) \over 1814400} $  &
$-{ \sqrt{2} t(3s^2-2su+3t^2)\over 15966720} $  &
$4su $  &
\tend
height12pt  & M17$({}^{--},{}^{+I},{}^{+J},{}^{IJ})$  &
$6 $  &
$0 $  &
$-{ t^2 \sqrt{2} \over 604800} $  &
$-{ t^3 \sqrt{2} \over 79833600} $  &
$4su $  &
\tend
height12pt  & M18$({}^{++},{}^{-I},{}^{+J},{}^{IJ})$ &
$6 $  &
$0 $  &
$-{ s(6s+7t) \sqrt{2} \over 1814400} $  &
${ s (10 t^2  + 26 t s + 11 s^2 )\sqrt{2}  \over 79833600} $  &
$ 2 {tu^2\over s}$  &
\tend
height12pt  & M19$({}^{++},{}^{++},{}^{IJ},{}^{IJ})$ &
$6 $  &
${tu \over 20160 s} $  &
${10t^2+21tu+10u^2 \over 453600} $  &
${s \, (20t^2 + 43tu+20u^2) \over 7983360} $  &
$4s^2 $  &
\tend
height12pt  & M20$({}^{--},{}^{++},{}^{IJ},{}^{IJ})$  &
$6 $  &
$0 $  &
${s^2 \over 453600} $  &
$-{s^3 \over 39916800} $  &
$4 { t^2u^2 \over s^2} $  &
\tend
height12pt  & M21$({}^{+I},{}^{+I},{}^{IJ},{}^{IJ})$  &
$6 $  &
${ tu \over 20160} $  &
${s (19 t^2  + 27 u t + 19 u^2 )\over 1209600 } $  &
${105 t^4  + 450 u t^3  + 671 u^2  t^2  + 450 u^3  t + 105 u^4
 \over 79833600 }$  &
$8s $  &
\tend
height12pt  & M22$({}^{+I},{}^{-I},{}^{IJ},{}^{IJ})$  &
$6 $  &
${s^2 \over 40320}  $  &
${ s (3 t^2  + 4 u t + 3 u^2)  \over 403200} $  &
${ s^2 (10 t^2  + 47 u t + 10 u^2   )\over   79833600}$  &
$8{tu \over s}  $  &
\tend
height12pt  & M23$({}^{+I},{}^{+I},{}^{JK},{}^{JK})$  &
$7 $  &
$\bullet $  &
${s \, (20t^2+33tu+20u^2) \over 1814400} $  &
${100t^4+398t^3u +593t^2u^2+398tu^3+100u^4 \over 79833600} $  &
$8s $  &
\tend
height12pt  & M24$({}^{+I},{}^{-I},{}^{JK},{}^{JK})$  &
$7 $  &
$\bullet $  &
${s \, (9t^2 +16tu+9u^2) \over 1814400} $  &
${s^2 \, (17t^2+38tu+17u^2 )\over 79833600} $  &
$8 {tu \over s} $  &
\tend
height12pt  & M25$({}^{+I},{}^{+J},{}^{IK},{}^{JK})$  &
$7 $  &
$\bullet $  &
$-{7t^3+16t^2u+16tu^2+10u^3 \over 3628800} $  &
${t \, (5t^3+26t^2u+39tu^2+26u^3) \over 79833600} $  &
$8s $  &
\tend
height12pt  & M26$({}^{+I},{}^{-J},{}^{IK},{}^{JK})$  &
$7 $  &
$\bullet $  &
${s \, (3t^2+2tu+6u^2) \over 3628800} $  &
${s(t^3 - u t^2  - u^2  t + 6 u^3 )\over 79833600 } $  &
$8{t u \over s}  $  &
\tend
height12pt  & M27$({}^{++},{}^{IJ},{}^{JK},{}^{KI})$  &
$7 $  &
$\bullet $  &
$0 $  &
$-{1 \over \sqrt{2}} \, {s^2 \, (s^2+t^2+u^2) \over 159667200} $  &
$8 {tu \over s} $  &
\tend
height12pt  & M28$({}^{IJ},{}^{IJ},{}^{IJ},{}^{IJ})$  &
$6 $  &
${ stu \over 26880} $  &
${11 (s^2+st+t^2)^2 \over 604800} $  &
${611 stu(s^2+st+t^2) \over 159667200 }  $  &
$16 $  &
\tend
height12pt  & M29$({}^{IJ},{}^{IJ},{}^{IK},{}^{IK})$  &
$7 $  &
$\bullet $  &
${ 19 t^4  + 56 u t^3  + 80 u^2  t^2  + 56 u^3  t + 19 u^4
\over 2419200 } $  &
${s(105 t^4  + 440 u^3 t  + 624 u^2  t^2  + 440 u^3  t + 105 u^4 )
\over 159667200} $  &
$16 $  &
\tend
height12pt  & M30$({}^{IJ},{}^{IJ},{}^{KL},{}^{KL})$  &
$8$  &
$\bullet $  &
${10t^4+32ut^3+45u^2t^2+32u^3t+10u^4\over 1814400 } $  &
${ s (100 (t^4+u^4)  + 381( u^3 t+ut^3)  + 555 u^2  t^2
 )\over 159667200 } $  &
$16 $  &
\tend
height12pt  & M31$({}^{IJ},{}^{JK},{}^{KL},{}^{LI})$  &
$8 $  &
$\bullet $  &
${(7s^4 +8s^3t+12s^2t^2+8st^3+7t^4) \over 7257600  } $  &
${ (s + t)
(5 (s^4+t^4)  - 12( s^3 t+st^3) - 2 s^2  t^2 )\over 159667200} $  &
$16 $  &
\tend
}
}
}
}
\nobreak
\caption[]{Infinities in the Graviton One-Loop Amplitudes due to a
Circulating Real Scalar
\label{GOLA}
\small
\smallskip}
\end{table}

\vfill\eject

\renewcommand{\theequation}{D.\arabic{equation}}
\renewcommand{\thetable}{D.\arabic{table}}
\renewcommand{\thesection}{D.\arabic{section}}
\renewcommand{\thesubsection}{D.\arabic{section}}
\setcounter{equation}{0}  
\setcounter{section}{1}  
\setcounter{table}{0}  

\section*{Appendix D: Infinities in One-Loop Graviton $2$-Form Amplitudes
and Four $2$-Form Amplitudes}

\twelverm

Here we present sufficient one-loop two graviton two $2$-form amplitudes
and four $2$-form amplitudes to determine the counterterms described in the main text.

The tables give both the tree amplitudes, which are of the form
$$
\CMtree(1,2,3,4) = i \, \kappa^2 \, K \times \Ftree
$$
and the infinities in the one-loop amplitudes
$$
\left. \CMloop(1,2,3,4) \right|_{1/\eps} =
 { i \kappa^4 \over (4\pi)^{D/2} \eps}
\times K
\times \Floop
$$
where $|K|$, $\Ftree$ and $\Floop$ are presented in tables~\ref{BTA} and~\ref{InfinitiesGGBB}.

\vskip 1cm

\begin{table}[!h]
\hbox{
\def\tend{\cr \noalign{ \hrule}}
\def\t#1{\tilde{#1}}
\def\tw{\theta_W}
\vbox{\offinterlineskip
{
\hrule
\halign{
       &  \vrule#
        &\strut\hfil #\hfil\vrule
        &\strut\hfil #\hfil\vrule
        &\strut\hfil #\hfil\vrule
        &\strut\hfil #\hfil\vrule
        &\strut\hfil #\hfil\vrule
        &\strut\hfil #\hfil\vrule
       \cr
height13pt  &{\bf Amplitude}  & $D_M$  & 
$\Floop, D=6$  & $\Floop, D=8$ & $\Ftree$ & $16 |K|$  &\tend
height13pt  & M$(1_B^{+I},2_B^{+I},3_B^{+I},4_B^{+I})$ &
$5 $  &
${t \over 2240} $  &
${ (s^2  + s t + t^2)^2 \over 20160su }  $  &
$0 $  &
$4su $  &
\tend
height13pt  & M$(1_B^{-I},2_B^{+I},3_B^{+I},4_B^{+I})$ &
$5 $  &
$0 $  &
$0 $  &
$0 $  &
$4tu $  &
\tend
height13pt  & M$(1_B^{-I},2_B^{-I},3_B^{+I},4_B^{+I})$ &
$5 $  &
$- {s \over 6720}$  &
$- {3 t^2  +2 u t + 3 u^2 \over 120960 } $  &
${ 2t^2+tu+2u^2 \over 2stu }  $  &
$4s^2 $  &
\tend
height13pt  & M$(1_B^{+I},2_B^{+I},3_B^{+J},4_B^{+J})$ &
$6 $  &
$-{(2t^2+tu+2u^2) \over 6720 u} $  &
${(t^2+tu+u^2)^2 \over 60480 su} $  &
$0$  &
$4su $  &
\tend
height13pt  & M$(1_B^{-I},2_B^{+I},3_B^{+J},4_B^{+J})$ &
$6 $  &
$0 $  &
$0 $  &
$0 $  &
$4tu  $  &
\tend
height13pt  & M$(1_B^{-I},2_B^{-I},3_B^{+J},4_B^{+J})$ &
$6 $  &
$-\frac{s}{6720} $  &
$ \frac{s^2}{120960} $  &
$ -{1 \over 2s} $  &
$4s^2  $  &
\tend
height13pt  & M$(1_B^{-I},2_B^{+I},3_B^{-J},4_B^{+J})$ &
$6 $  &
${(s-t)\over 6720} $  &
${ t^2  + 3 s^2  + 2 s t \over 120960}$  &
${(t-u)\over 2su } $  &
$4u^2  $  &
\tend
height13pt  & M$(1_B^{+I},2_B^{+I},3_B^{IJ},4_B^{IJ})$ &
$6 $  &
$-{ t^2+u^2 \over 13440} $  &
${ s(t^2  + t u + u^2 )\over 80640} $  &
$0 $  &
$8s $  &
\tend
height13pt  & M$(1_B^{+I},2_B^{-I},3_B^{IJ},4_B^{IJ})$ &
$6 $  &
${s^2 \over 13440} $  &
$  { s(3 t^2  + 4 t u + 3 u^2 )\over 241920}$  &
${ (t^2+u^2 )\over 4tu } $  &
$8{tu \over s}  $  &
\tend
height13pt  & M$(1_B^{+I},2_B^{+I},3_B^{JK},4_B^{JK})$ &
$7 $  &
$\bullet $  &
$0 $  &
$0 $  &
$8s  $  &
\tend
height13pt  & M$(1_B^{+I},2_B^{-I},3_B^{JK},4_B^{JK})$ &
$7 $  &
$\bullet $  &
${s^3 \over 120960} $  &
$-{1\over 2} $  &
$8{tu \over s}  $  &
\tend
height13pt  & M$(1_B^{+I},2_B^{+J},3_B^{IK},4_B^{JK})$ &
$7 $  &
$\bullet $  &
${(2s+t) \, (s^2+st+t^2) \over 241920} $  &
$0$  &
$8s $  &
\tend
height13pt  & M$(1_B^{+I},2_B^{-J},3_B^{IK},4_B^{JK})$ &
$7 $  &
$\bullet $  &
${su^2 \over 241920} $  &
$-{s \over 4u}  $  &
$8{tu \over s} $  &
\tend
height13pt  & M$(1_B^{IJ},2_B^{IJ},3_B^{IJ},4_B^{IJ})$ &
$6 $  &
$0 $  &
${ (u^2  + t u + t^2 )^2 \over 60480}$  &
${(s^2+t^2+u^2)^2 \over 16stu} $  &
$16 $  &
\tend
height13pt  & M$(1_B^{IJ},2_B^{IJ},3_B^{IK},4_B^{IK})$ &
$7 $  &
$\bullet $  &
${ (3 t^4  + 6 u t^3  + 8 u^2  t^2  + 6 u^3  t + 3 u^4)\over 483840} $  &
$-{t^2+u^2 \over 8s} $  &
$16 $  &
\tend
height14pt  & M$(1_B^{IJ},2_{B}^{IJ},3_B^{KL},4_B^{KL})$ &
$8 $  &
$\bullet $  &
$-{s^2tu \over 241920} $  &
${tu \over 4s} $  &
$16 $  &
\tend
height13pt  & M$(1_B^{IJ},2_B^{JK},3_B^{KL},4_B^{LI})$ &
$8 $  &
$\bullet $  &
${s^4+t^4 \over 483840} $  &
${u\over 8}$  &
$16 $  &
\tend
}
}
}
}
\nobreak
\caption[]{The Tree Amplitudes and One-Loop Infinities for the Four $2$-Form Amplitudes
\label{BTA}
\small
\smallskip}
\end{table}

\vfill\eject


\begin{table}[!t]
\hbox{
\def\tend{\cr \noalign{ \hrule}}
\def\t#1{\tilde{#1}}
\def\tw{\theta_W}
\vbox{\offinterlineskip
{
\hrule
\halign{
       &  \vrule#
        &\strut\hfil #\hfil\vrule
        &\strut\hfil #\hfil\vrule
        &\strut\hfil #\hfil\vrule
        &\strut\hfil #\hfil\vrule
        &\strut\hfil #\hfil\vrule
        &\strut\hfil #\hfil\vrule
       \cr
height12pt  &{\bf Amplitude}  & $\Floop, D=6$   & $\Floop,D=8$
& $\Ftree$  & $16|K|$ &\tend
height12pt  & M$({g}^{+},{g}^{+},{B}^{+I},{B}^{+I})$ &
 $0 $   &
${ s^2 \over 25200 u}$  &
$0$   & ${2su^2 }$  &  \tend
height12pt  & M$({g}^{-},{g}^{-},{B}^{+I},{B}^{+I})$ &
 $0 $   &
${s \over 25200} $  &
$0$   & ${2s^3}$  & \tend
height12pt  & M$({g}^{-},{g}^{+},{B}^{+I},{B}^{+I})$ &
 $0 $   &
$0$  &
$0 $   & ${2su^2}$  &  \tend
height12pt  & M$({g}^{+},{g}^{+},{B}^{+I},{B}^{-I})$ &
 $ - { t\over 10080 s}$   &
$- { t \over 100800}$  &
$0 $   & ${2s^2u}$  &  \tend
height12pt  & M$({g}^{-},{g}^{+},{B}^{-I},{B}^{+I})$ &
 $0 $   &
$-{ t \over 50400 } $  &
$-{ 2 \over su } $   & ${2u^3}$  & \tend
height12pt  & M$({g}^{+},{g}^{+},{B}^{IJ},{B}^{IJ})$ &
 $-{ 3 t^2  + 5 u t + 3 u^2 \over 20160 s }$   &
${ 2 t^2  + 5 u t + 2 u^2 \over  201600}$  &
$0$   & ${4s^2}$  & \tend
height12pt  & M$({g}^{+I},{g}^{+I},{B}^{+J},{B}^{+J})$ &
 $-{ s^2 \over 13440  u}  $   &
${ s^3 \over 67200 u }$  &
$0  $   & ${4su}$  &   \tend
%
%
height12pt  & M$({g}^{-I},{g}^{+I},{B}^{+J},{B}^{+J})$  &
$0$   &
$0$  &
$0$   &
$4tu$ &   \tend
height12pt  & M$({g}^{+I},{g}^{+J},{B}^{-I},{B}^{+J})$  &
$0 $   &
$0$  &
$0$   & 
$4st$ 
&   \tend
%
%
height12pt  & M$({g}^{-I},{g}^{+J},{B}^{+I},{B}^{+J})$  &
$0$   &
$0$  &
$0$   &
$4st$  &    \tend
%
%
height12pt  & M$({g}^{-I},{g}^{-I},{B}^{+J},{B}^{+J})$  &
$- {s \over 13440} $   &
${s^2 \over 67200} $   &
$-{1 \over 2s } $   & 
$4s^2$ &    \tend
%
%
height12pt  & M$({g}^{-I},{g}^{+J},{B}^{-I},{B}^{+J})$  &
$0 $  &
$0 $ &
$-{1 \over 2u }$ &
$4u^2$    &    \tend
%
%
height12pt  & M$({g}^{-I},{g}^{+I},{B}^{-J},{B}^{+J})$  &
${s \over 6720 } $  &
${s^2 \over  100800 }$ &
${ t-u \over 2su  }  $  &
$4u^2$ &    \tend
%
%
height12pt  & M$({g}^{+},{g}^{+I},{B}^{+J},{B}^{IJ})$  &
${s \sqrt{2} \over 13440} $  &
${s^2 \sqrt{2} \over 201600} $ &
$0 $  &
$4su$  &    \tend
%
%
height12pt  & M$({g}^{+I},{g}^{+I},{B}^{JK},{B}^{JK})$  &
$\bullet$  &
${ s  (2\,{t}^{2}+3\,ut+2\,{u}^{2} )
\over 403200}$ &
$0$  &
$8s$ &    \tend
%
%
height12pt  & M$({g}^{+I},{g}^{+I},{B}^{IJ},{B}^{IJ})$  &
$-{tu \over 20160} $  &
${ s (4 u^2  + 7 t u + 4 t^2 )  \over 403200}$ &
$0$  &
$8s$ &    \tend
%
%
height12pt  & M$({g}^{JK},{g}^{JK},{B}^{+I},{B}^{+I})$  &
$\bullet $  &
${\frac {1}{201600}}\,{s}^{3} $ &
$0$  &
$8s$ &    \tend
%
%
height12pt  & M$({g}^{--},{g}^{+I},{B}^{+J},{B}^{IJ})$   &
$0$   &
$0$  &
$-{ \sqrt{2} t \over 2 s u}$   & 
$4su$ &    \tend
height12pt  & M$({g}^{+I},{g}^{-I},{B}^{JK},{B}^{JK})$   &
$\bullet$   &
${s^3 \over 201600} $  &
$-{1 \over 2} $   &
$8{tu \over s}$ &    \tend
height12pt  & M$({g}^{+I},{g}^{-I},{B}^{IJ},{B}^{IJ})$   &
${s^2 \over  13440}$   &
${s^3 \over  201600}$  &
${ (t^2+u^2) \over 4tu } $   &
$8{tu \over s}$ &    \tend
height12pt  & M$({g}^{JK},{g}^{JK},{B}^{+I},{B}^{-I})$   &
$\bullet $   &
${s^3 \over 134400} $  &
$-{1 \over 2}$   & 
$8{tu \over s}$&    \tend
height12pt  & M$({g}^{+},{g}^{-I},{B}^{+J},{B}^{IJ})$   &
$0  $   &
$0 $  &
$0$   &
$2{tu^2 \over s}$  &    \tend
%
height12pt  & M$({g}^{+},{g}^{+I},{B}^{-J},{B}^{IJ})$   &
${u \sqrt{2}  \over 13440} $   &
${s u \sqrt{2}  \over201600} $  &
$0$   &
$2{ts^2 \over u}$  &    \tend
%
height12pt  & M$({g}^{+I},{g}^{+J},{B}^{IK},{B}^{JK})$   &
$\bullet$   &
${s^3 \over 403200} $  &
${0}$   & 
$8s$ &    \tend
%
height12pt  & M$({g}^{+I},{g}^{IK},{B}^{+J},{B}^{JK})$   &
$\bullet $   &
${s^3 \over 403200} $  &
$0$   &
$8u$ 
 &    \tend
%
height12pt  & M$({g}^{IK},{g}^{JK}.{B}^{+I},{B}^{+J})$   &
$\bullet$   &
${s^3 \over 403200}$  &
$0$   &
$8s $ &    \tend
%
height12pt  & M$({g}^{IJ},{g}^{IJ}.{B}^{+I},{B}^{+I})$   &
$0$   &
${ s^3 \over 100800} $  &
$0$   &
$8s $ &    \tend
%
height12pt  & M$({g}^{IJ},{g}^{JK},{B}^{KL},{B}^{LI})$ &
$\bullet$   &
$-{\frac {1}{806400}}\,{s}^{4}$  &
$-{u \over 8}$
&$16$ &    \tend
height12pt  & M$({g}^{--},{g}^{++},{B}^{IJ},{B}^{IJ})$ &
$0$   &
$0$  &
${s \over tu } $    & 
$4{ t^2u^2 \over s^2 }$ &  \tend
%
height12pt  & M$({g}^{IJ},{g}^{KL},{B}^{JK},{B}^{LI})$ &
$\bullet$   &
$0$  &
$-{s \over 8}$    & 
$16$ &  \tend
%
height12pt  & M$({g}^{IJ},{g}^{IJ},{B}^{KL},{B}^{KL})$   &
$\bullet$   &
$ {(2\,{t}^{2}+ut+2\,{u}^{2} ){s}^{2} \over 806400} $  &
${tu\over 4s} $   &
$16$  &  \tend
height12pt  & M$({g}^{IJ},{g}^{KL},{B}^{IJ},{B}^{KL})$  &
$\bullet$   &
$0$  &
${ts\over 4u}$    & 
$16$ &  \tend
}
}
}
}
\nobreak
\caption[]{The Tree Amplitudes and One-Loop Infinities for the
Two Graviton Two $2$-Form Amplitudes
\label{InfinitiesGGBB}
\small
\smallskip}
\end{table}

\vskip 1.0 truecm 

\null

\vfill\eject


\end{document}